\documentclass[12pt]{iopart}

\usepackage{iopams}  
\usepackage{pifont}
\usepackage[hyperindex,breaklinks]{hyperref}
\expandafter\let\csname equation*\endcsname\relax
\expandafter\let\csname endequation*\endcsname\relax
\usepackage[caption=false]{subfig}
\usepackage{amsmath}
\usepackage{amssymb}
\usepackage{graphicx}
\usepackage{tensor}
\usepackage{setspace}
\usepackage{pzccal}
\usepackage[sort&compress,numbers]{natbib}
\usepackage{graphicx}
\usepackage{epstopdf}
\usepackage{appendix}
\usepackage{alphalph}
\usepackage{manfnt} 
\usepackage{tikz}
\usetikzlibrary{arrows,positioning,automata,shadows,fit,shapes}
\usetikzlibrary{fit}
\usetikzlibrary{decorations.pathreplacing}
\usepackage{color}
\setcitestyle{square,comma}
\usepackage[symbol*]{footmisc}

\def\c#1{\overline{#1}}

\definecolor{myblue}{rgb}{0.1, 0.1, 0.8}
\definecolor{myred}{rgb}{0.7, 0.1, 0.1}
\definecolor{mygreen}{rgb}{0, 0.5, 0}

\newcommand{\one}{\text{\ding{172}}}
\newcommand{\two}{\text{\ding{173}}}
\newcommand{\three}{\text{\ding{174}}}
\newcommand{\One}{\text{\ding{202}}}
\newcommand{\Two}{\text{\ding{203}}}
\newcommand{\Three}{\text{\ding{204}}}
\newcommand{\Four}{\text{\ding{205}}}
\newcommand{\none}{\rm{n}^\one }
\newcommand{\ntwo}{\rm{n}^\two }
\newcommand{\fone}{\rm{f}^\one \,}
\newcommand{\ftwo}{\rm{f}^\two \,}
\newcommand{\Rone}{$\rm{R}_\one \,$}
\newcommand{\Rtwo}{$\rm{R}_\two \,$}
\newcommand{\Rthree}{$\rm{R}_\three \,$}
\newcommand{\Tone}{\rm{\mathbf{T}}^\one}
\newcommand{\Ttwo}{\rm{\mathbf{T}}^\two}
\newcommand{\Tthree}{\rm{\mathbf{T}}^\three}
\newcommand{\Aone}{{\rm{\mathbf{A}}^\one}}
\newcommand{\Atwo}{{\rm{\mathbf{A}}^\two}}
\newcommand{\Athree}{{\rm{\mathbf{A}}^\three}}
\newcommand{\Bone}{{\rm{\mathbf{B}}^\one}}
\newcommand{\Btwo}{{\rm{\mathbf{B}}^\two}}
\newcommand{\Bthree}{{\rm{\mathbf{B}}^\three}}
\newcommand{\Cone}{{\rm{\mathbf{C}}^\one}}
\newcommand{\Ctwo}{{\rm{\mathbf{C}}^\two}}
\newcommand{\Cthree}{{\rm{\mathbf{C}}^\three}}
\newcommand{\Done}{{\rm{\mathbf{D}}^\one}}
\newcommand{\Dtwo}{{\rm{\mathbf{D}}^\two}}
\newcommand{\Dthree}{{\rm{\mathbf{D}}^\three}}
\begin{document}
\title[Reduced phase space optics for general relativity: Symplectic ray bundle transfer]{Reduced phase space optics for general relativity: Symplectic ray bundle transfer}

\author{Nezihe Uzun}

\address{Institute of Theoretical Physics, Faculty of Mathematics and Physics, Charles University in Prague, Prague, Czech Republic}
\address{Univ Lyon, Ens de Lyon, Univ Lyon1, CNRS, Centre de Recherche Astrophysique de Lyon UMR5574, F–69007, Lyon, France }
\vspace{10pt}

\begin{indented}
\item[]10 June 2019
\end{indented}

\begin{abstract}
In the paraxial regime of Newtonian optics, propagation of an ensemble of rays is represented by a symplectic ABCD transfer matrix defined on a reduced phase space. Here, we present its analogue for general relativity. Starting from simultaneously applied null geodesic actions for two curves, we obtain a geodesic deviation action up to quadratic order. We achieve this by following a preexisting method constructed via Synge's world function. We find the corresponding Hamiltonian function and the reduced phase space coordinates that are composed of the components of the Jacobi fields projected on an observational screen. Our thin ray bundle transfer matrix is then obtained through the matrix representation of the Lie operator associated with this quadratic Hamiltonian. Moreover, Etherington's distance reciprocity between any two points is shown to be equivalent to the symplecticity conditions of our ray bundle transfer matrix. We further interpret the bundle propagation as a free canonical transformation with a generating function that is equal to the geodesic deviation action. We present it in the form of matrix inner products. A phase space distribution function and the associated Liouville equation is also provided. Finally, we briefly sketch the potential applications of our construction. Those include reduced phase space and null bundle averaging; factorization of light propagation in any spacetime uniquely into its thin lens, pure magnifier and fractional Fourier transformer components; wavization of the ray bundle; reduced polarization optics and autonomization of the bundle propagation on the phase space to find its invariants and obtain the stability analysis.
\end{abstract}
\maketitle

\section{Introduction}
Reciprocity relations in physics signal the existence of potentiality of a system \citep{Tulczyjew:1977,Kijowski:1979}. Maxwell-Betti reciprocity for virtual work in elasticity \citep{Yao:2009}, Onsager's reciprocity in thermodynamics \citep{Onsager:1931} or quantum mechanical reciprocity of the received signal \citep{Bilhorn:1964} all state that the observables are unchanged when the input and output agents are traversed. Those distinct systems share a similar property: they can be linked to some well-defined symplectic potential. The work we present here grew out of questioning what kind of potentiality Etherington's distance reciprocity in relativity \citep{Etherington:1933} corresponds to. The outcome of such an investigation turns out to be a symplectic phase space reformulation of first order geometric optics in relativity. 

Observationally viable studies of optics in general relativity are usually investigated under two main branches: i) gravitational lensing studies, ii) cosmological light propagation. For gravitational lensing calculations, one chooses an approximate stationary metric and an appropriate $3+1$ decomposition of the spacetime. The equations of object and image distances can be derived all the way from Fermat's principle. Then, potentials and refractive indices analogous to the ones of the Newtonian theory can be obtained \citep{Schneider:1999}. On the other hand, such analogies between the Newtonian and the general relativistic Fermat's principle cannot be formed for distance calculations in cosmology as the underlying metric is far from being stationary. For instance, angular diameter and luminosity distances can be obtained via the Jacobi fields \citep{Ellis:1971} whose relation to an analogue refractive index is not clear. 

Our aim here is to propose a method in order to study the phase space propagation of a thin ray bundle defined within any spacetime. While doing this, we consider a reduced phase space, as in the case of paraxial regime of Newtonian optics, such that the propagation of the bundle within a spacetime and a classical optical device are analogous up to first order. Therefore, in Section~\ref{First order Newtonian ray optics}, we summarize the paraxial ray optics of the Newtonian theory and remind how symplectic ray transfer matrices emerge on a reduced phase space. Most of the notation used in our construction is introduced there. In Section~\ref{First order phase space ray optics for curved background}, main ideas behind our work are presented. We start by the application of geodesic actions simultaneously for two null curves. Then following the method of  \citep{Vines:2014} which involves a bilocal function known as the Synge's world function \citep{Synge:1960}, we obtain a geodesic deviation action up to quadratic order. This is applicable for nearly parallel, neighboring null geodesics and hence analogous to the Newtonian paraxial regime. A corresponding Hamiltonian formalism for a 4-dimensional phase space is obtained once we consider the observational screen projections of the Jacobi fields as phase space coordinates. Hence, symplectic ray bundle transfer matrices are constructed. In Section~\ref{Distances, reciprocity and symplecticity}, we show that Etherington's distance reciprocity indeed follows from the symplecticity conditions of this transfer matrix. In Section~\ref{Canonical transformations and generating functions}, we provide the generating function of the linear canonical transformation corresponding to the symplectomorphism of our phase space. Moreover, a phase space distribution function for the ensemble of rays  and its corresponding Liouville's equation is provided in Section~\ref{Density function and Liouville's equation}. In the end, in Section~\ref{Potential applications of reduced phase space optics}, we propose certain potential applications of symplectic ray bundle transfer matrices for astrophysical and cosmological scenarios. These include: (i) phase space and null bundle averaging of scalars which can then be used to average Einstein equations; (ii) factorizing the light propagation effect in any spacetime into its thin lens, pure magnifier and fractional Fourier transformer components; (iii) wavization of a ray bundle; (iv) investigating the evolution of polarization states and (v) determining the invariants and stability analysis of a null bundle by considering some autonomization techniques. The last Section~\ref{Summary and conclusion} gives a summary and conclusion of the work.

We choose the $(-, +, +, +)$ signature for our spacetime metric and also use natural units through out the paper so that $c, G, h, k_B$ are set to $1$.
\section{First order Newtonian ray optics}\label{First order Newtonian ray optics}
 {In this section, we will remind the first order Newtonian optics by giving a brief summary of references \citep{Holm:2008, Torre:2005,Wolf:2004}. Our aim is to re-present the reader how symplectic transfer matrices emerge from Fermat's principle, as a similar construction for general relativity will be developed later. Even though we claim no new results in this section, note that most of the notation and concepts relevant for the paper is introduced here.}

\subsection{Fermat's principle and paraxial approximation}\label{Fermat's principle and paraxial approximation}
Let us consider a stationary, inhomogeneous and isotropic medium. According to Fermat's principle, the path of a ray is the one that extremizes the following action between points $P_1$ and $P_2$
\begin{eqnarray}\label{eq:Action_Newtonian}
A=\int_{P_1}^{P_2}n(\vec{r})\,ds.
\end{eqnarray}
Here, $\vec{r}(s)\in \mathbb{R}^3$ is the position vector, $n(\vec{r})=c/V$ is the refractive index of the medium with $c$ being the speed of light in vacuum and $V$, the one in the medium. The Euclidean arc length is denoted by $ds^2=d\vec{r}\cdot d\vec{r}$.

In order to obtain the \textit{eikonal equation}, i.e., the equations of motion, one is free to pick more than one parameterization and/or degree of the Lagrangian function associated with the action $A$. In the literature, however, one of the common approaches is to write the  equations of motion, with respect to the Euclidean arc length, so that the solution of
\begin{eqnarray}\label{FermatL}
\delta A=\int_{P_1}^{P_2}\delta\left(n(\vec{r})\sqrt{\dot{\vec{r}}\cdot \dot{\vec{r}}}\right)ds=0,
\end{eqnarray}
gives the equations of motion
\begin{eqnarray}
\frac{d}{ds}\left(\frac{\partial \tilde{L}}{\partial \dot{\vec{r}}}\right)-\frac{\partial \tilde{L}}{\partial\vec{r}}=0,
\end{eqnarray} 
with
\begin{eqnarray}\label{eq:L(r,dotr)}
\tilde{L}=n(\vec{r})\sqrt{\dot{\vec{r}}\cdot \dot{\vec{r}}}
\end{eqnarray}
being the Lagrangian function and the overdot denotes a total derivative with respect to the arc length $s$. Note that we have the normalization $|\dot{\vec{r}}|=1$ here and the Lagrangian is a homogeneous function of degree one with respect to $\dot{\vec{r}}$. Then the eikonal equation is written as
\begin{eqnarray}\label{eq:eikonal}
\frac{d}{ds}\left[n\left(\vec{r}\right)\cdot\frac{d\vec{r}}{ds}\right]=\nabla n\left(\vec{r}\right),
\end{eqnarray}
where $\nabla$ is the gradient operator defined with respect to the Euclidean metric.

One can switch from the Lagrangian formulation to a Hamiltonian formulation, by considering $\{\vec{r}, \dot{\vec{r}}\}$ as the canonical coordinates and velocities respectively, so that
\begin{eqnarray} 
\tilde{H}=\frac{\partial \tilde{L}\left(\vec{r}, \dot{\vec{r}}\right)}{\partial\dot{\vec{r}}}\cdot\dot{\vec{r}}-\tilde{L}=\vec{p}\cdot\dot{\vec{r}}-\tilde{L}.
\end{eqnarray}
{Note that this is the total, conserved Hamiltonian of the system which is equivalent to $\tilde{H}=|\vec{p}|-n=0$ on shell.}

In order to get to the paraxial approximation, it is a common practice to start the procedure by reparameterizing the optical equations with respect to one of the configuration space coordinates. Let us pick it to be the  {$y-$}coordinate of the Euclidean distance $ds= {\left(du^2 + dv^2 + dy^2\right)^{1/2}=\beta dy}$ with  {$\beta =\left(u^{\prime 2}+v^{\prime 2}+1\right)^{1/2}$}. Here, prime denotes the total derivative with respect to the new evolution parameter  {$y$}. Then Fermat's action in (\ref{FermatL}) can be recast in the following form \citep{Holm:2008}
\begin{eqnarray}\label{eq:FermatredL}
A&=&\int_{P_1}^{P_2}  {L(\mathbf{q},\mathbf{q}^\prime ;y)\,dy}\\ \nonumber
&=&\int_{P_1}^{P_2}  {n(\mathbf{q}; y)}\left(1+|\mathbf{q}^\prime|^2\right)^{1/2}\, {dy},
\end{eqnarray}
if we pick our optical canonical coordinates via $d\mathbf{q}= {(du,dv)}$ that lies on a screen orthogonal to some optical axis for each value of  {$y$}. The Lagrangian function defined in (\ref{eq:FermatredL}) is often referred to as the \textit{optical/reduced/screen Lagrangian}. 

Once we apply a Legendre transformation on the reduced Lagrangian ${L}$ we get the reduced Hamiltonian
\begin{eqnarray}\label{eq:redHNewton}
H=\mathbf{p}\cdot\mathbf{q}^\prime-L=-\left[ {n(\mathbf{q};y)}^2-|\mathbf{p}|^2\right]^{1/2},
\end{eqnarray}
where $\mathbf{p}=\left(n/\beta\right)\mathbf{q}^\prime$ and the Hamilton-Jacobi equations read as
\begin{eqnarray}\label{eq:red_Hamilton-Jacobi}
\mathbf{q}^\prime=\frac{ {\partial} H}{ {\partial} \mathbf{p}},\qquad \mathbf{p}^\prime=-\frac{ {\partial} H}{ {\partial} \mathbf{q}}.
\end{eqnarray} 
Note that unlike the total Hamiltonian $\tilde{H}$, the reduced Hamiltonian $H$ is not conserved throughout the evolution.

In the paraxial approximation, the angle $\Delta \theta$ between the propagation vector of light and the optical axis is assumed to be small\footnote{For practical purposes, angles smaller than 15 degrees are well within this approximation in the Newtonian theory.}, i.e.,  {$ds\approx dy$} or $\beta \approx 1$. In that case, the optical momentum $\mathbf{p}$ is a measure of the angle in question, i.e., $ {|}\mathbf{p} {|}=n\Delta \mathbf{\theta}$. Again, it is a common practice to Taylor expand the reduced Hamiltonian, (\ref{eq:redHNewton}), with respect to the optical momentum only and obtain the Hamiltonian for the first order ray propagation as \citep{Torre:2005, Wolf:2004}
\begin{eqnarray}\label{eq:HTaylor_p}
H=-\left(n-\frac{|\mathbf{p}|^2}{2n}-\frac{|\mathbf{p}|^4}{8n^3}-\frac{|\mathbf{p}|^6}{16n^5}- ...\right)\approx \frac{|\mathbf{p}|^2}{2n}-n.
\end{eqnarray}

The so-called ABCD ray transfer matrices in optics are very much related to \textit{quadratic} Hamiltonians. Those transfer matrices take the optical system from one set of solutions, $(\mathbf{q},\mathbf{p})$, to another one, $(\mathbf{Q},\mathbf{P})$. In order to obtain them, one further expands the refractive index $ {n(\mathbf{q};y)}$ around its value on the optical axis, i.e., at $ {(\mathbf{0};y)}$, up to quadratic order. We will show this in the next section. 
\subsection{Symplectic geometry and ABCD matrices}\label{Symplectic geometry and ABCD matrices}
Let us expand (\ref{eq:HTaylor_p}) with respect to the canonical coordinates $\mathbf{q}$ of a centered system, for instance. Then, the first order terms vanish, as they represent the tilts and misalignments with respect to the optical axis. Moreover, the zeroth order term will not be essential when we introduce the Lie operator and thus we omit it. Then one rewrites the Hamiltonian (\ref{eq:HTaylor_p}) as \citep{Torre:2005}
\begin{eqnarray}\label{eq:Hquad}
H&=&\frac{1}{2n_0}\delta ^{ab}p_ap_b-\frac{1}{2}n_{ab}q^aq^b,
\end{eqnarray}
where  {$\{a,b\}=\{\hat{u}, \hat{v}\}$}, $\delta _{ab}$ is the Kronecker delta function in 2-dimensions,  {$n_0=n(\mathbf{0};y)$} and  {$n_{ab}(\mathbf{0};y)$} represents the second order variation of the refractive index with respect to the canonical coordinates.

Note that the Hamiltonian in (\ref{eq:Hquad}) is quadratic with respect to both $p_a$'s and $q^a$'s. Those polynomials are very important in many areas of physics as they are closed under the Poisson bracket and thus form a Lie algebra. Our aim here is to introduce the Lie operator corresponding to the reduced Hamiltonian. Its matrix representation is a Hamiltonian block matrix that evolves the first order system in question.

In order to show this, let us introduce a $2n$ dimensional symplectic phase space $M(\mathbb{R}^{2n})$. We will denote the phase space coordinates as $z^i= {(q^a, \,p_b)^{\intercal}}$ where $\{a, b\}=\{1...n\}$, $\{i, j\}=\{1...2n\}$ and $^{\intercal}$ refers to the transpose operator. In the current section $n=2$, however, the following construction is valid for any dimensions. 

Poisson bracket of two functions $f$ and $g$ is given by
\begin{eqnarray}\label{eq:Poisson_funcs}
\{f,g\}=\frac{\partial f}{\partial z^i}\Omega ^{ij}\frac{\partial g}{\partial z^j},
\end{eqnarray}
where $\mathbf{\Omega}$ is the fundamental symplectic matrix\footnote{In the literature, $\mathbf{\Omega}$ is sometimes denoted as $\mathbf{J}$ or $\boldsymbol{\omega}$. The reader should also be careful about the sign convention chosen here.} defined through
\begin{eqnarray}\label{eq:Poisson_z}
\{z^i,z^j\}=\Omega ^{ij}, \qquad
\Omega ^{ij}=
\left[
\begin{array}{c|c}
\mathbf{0_n} & \, \, \mathbf{I_n} \\
\hline
\mathbf{-I_n} & \, \, \mathbf{0_n}
\end{array}
\right],
\end{eqnarray}
where $\mathbf{I_n}$ and $\mathbf{0_n}$ are identity and zero matrices, respectively, of dimension $n$. The matrix $\mathbf{\Omega}$ has the following properties
\footnote{
With lowered indices components of $\mathbf{\Omega}$ follows as
$
\Omega _{ij}=
\left[
\begin{array}{c|c}
\mathbf{0_n} \, \,& \mathbf{-I_n} \\
\hline
\mathbf{I_n} \, \,& \mathbf{0_n}
\end{array}
\right].
$
}
\begin{eqnarray}\label{eq:Omega_prop}
\mathbf{\Omega}^{\intercal}=\mathbf{\Omega}^{-1}=-\mathbf{\Omega},\,\,\,  \mathbf{\Omega}^2=-\mathbf{I_{2n}},\,\,\, \rm{det}\,\mathbf{\Omega}=1,
\end{eqnarray}
in which $^{-1}$ denotes the inverse operator and ${\rm{det}}$ refers to determinant of the matrix.
With this notation Hamilton-Jacobi equations (\ref{eq:red_Hamilton-Jacobi}) can be recast into
\begin{eqnarray}\label{eq:Hamilton_eqs_z}
\frac{dz^i}{ {dy}}=\Omega ^{ij}\frac{\partial H}{\partial z^j}=-\{H,z^i\}.
\end{eqnarray}
Let us denote the Lie operator corresponding to the Hamiltonian (\ref{eq:Hquad}) as
\begin{eqnarray}\label{eq:Lie_op_Newtonian}
\hat{\mathcal{L}}_{H}\left[\bullet \right]=-\{H,\bullet \}= \frac{1}{n_0}\delta ^{ab}p_b\frac{\partial}{\partial q^a}+n_{ab}q^b\frac{\partial}{\partial p_a}.
\end{eqnarray}
Note that since, $\hat{\mathcal{L}}_{H}$ is a Lie operator associated with a quadratic polynomial on a $2n$-dimensional phase space, there exists a $2n\times 2n$ matrix representation of it \citep{Srivastava:1984, Dattoli:1990}. We will denote it as 
\begin{eqnarray}\label{eq:Lie_matrix_Newtonian}
\mathbf{L}_{\mathbf{H}}=
\left[
\begin{array}{c|c}
\mathbf{0} & \, \, \mathbf{n_0}^{-1} \\
\hline
\mathbf{n_2} & \, \, \mathbf{0}
\end{array}
\right],
\end{eqnarray}
in which $\mathbf{n_2}$ and $\mathbf{n_0}$ have the components $n_{ab}$ and  $\delta _{ab}n_0$ respectively.
Then, for this linear system, (\ref{eq:Hamilton_eqs_z}) can be rewritten as
\begin{eqnarray}\label{eq:Hamilton_eqs_matrix}
\frac{d\mathbf{z}}{ {dy}}=\mathbf{L}_{\mathbf{H}}\,\mathbf{z}.
\end{eqnarray}
Let us consider the simplest case for now and assume that the refractive index is  {$y$}-independent. Then the evolution of the system between any initial and arbitrary points is given by
\begin{eqnarray}\label{eq:T}
\mathbf{z}=\mathbf{T} {\left(y,y_0\right)}\mathbf{z_0},
\end{eqnarray}
with
\begin{eqnarray}
\mathbf{T} {\left(y,y_0\right)}=\exp{[{\mathbf{L}_{\mathbf{H}} {\left(y,y_0\right)}}]}=\sum _{m=0}^{\infty} \frac{ {\left(y-y_0\right)}^m}{m!}\mathbf{L}_{\mathbf{H}}^m,
\end{eqnarray}
such that $\mathbf{T}$ represents a Lie transformation.

As exponential maps of Hamiltonian matrices are symplectic matrices \citep{deGosson:2011}, the ray transfer equation (\ref{eq:T}) is a linear symplectic transformation that preserves the Poisson bracket structure (\ref{eq:Poisson_z}). Then the symplectic matrix $\mathbf{T}$ satisfies
\begin{eqnarray}\label{eq:Symplectic_T}
\mathbf{T}^{\intercal}\,\mathbf{\Omega}\,\mathbf{T}=\mathbf{\Omega}, \qquad \rm{det}\,\mathbf{T}=1.
\end{eqnarray}
Note that it can be put in a block form
\begin{eqnarray}\label{eq:T_block}
\mathbf{T}=
\left[
\begin{array}{c|c}
\mathbf{A} & \mathbf{B} \\
\hline
\mathbf{C} & \mathbf{D}
\end{array}
\right],
\end{eqnarray}
with $\mathbf{A}, \mathbf{B}, \mathbf{C}$ and $\mathbf{D}$ being all $n$-dimensional square matrices. That is why $\mathbf{T}$ is usually referred to as an $ABCD$ \textit{matrix} in the literature. 

Now, let us substitute (\ref{eq:T}) back in (\ref{eq:Hamilton_eqs_matrix}). Then we obtain
\begin{eqnarray}\label{eq:T_evol}
\frac{d\mathbf{T}}{ {dy}}=\mathbf{L}_{\mathbf{H}}\,\mathbf{T},
\end{eqnarray}
as the initial phase space vector is fixed. Also, for the convenience of the next section, let us pick  {a} symmetric optical system so that, the refractive index is also  {$v$}-independent. Then matrices $\mathbf{n_2}$ and $\mathbf{n_0}$ reduce to scalars $n_2$ and $n_0$\footnote{\label{fn:GRIN}Note that, previously, when we Taylor expanded the refractive index, we actually assumed that it varies within the medium smoothly. That is why the corresponding medium is usually referred to as \textit{graded index} (GRIN) medium in the literature. For the design of optical instruments or fibers, researchers often assume an elliptic profile for the refractive index in which the second order term $n_2$ plays the major role in its identification.  {For more details cf. Chapters 2 and 3 of \citep{Gomez:2002} or Section 5.3 of \citep{Merchand:2012}.}  We believe $n_2$ having such a major role will be more clear in Section~\ref{First order phase space ray optics for curved background}, within the differential geometric language.}. Likewise the set $\{\mathbf{A}, \mathbf{B}, \mathbf{C},\mathbf{D}\}$ reduces to a set of scalars $\{A, B, C, D\}$ which define the transfer matrix of a $2$-dimensional phase space vector. In that case (\ref{eq:T_evol}) can be cast into a set of four first order differential equations  {\citep{Torre:2005, Wolf:2004}}
\begin{eqnarray}\label{eq:evol_ABCD}
\frac{dA}{ {dy}}&=&\frac{C}{n_0}, \qquad \frac{dB}{ {dy}}=\frac{D}{n_0}\\ \nonumber
\frac{dC}{ {dy}}&=&n_2 A, \qquad \frac{dD}{ {dy}}=n_2B,
\end{eqnarray}
with initial conditions
\begin{eqnarray}
A( {y_0})&=&1, \qquad B( {y_0})=0,\\ \nonumber
C( {y_0})&=&0, \qquad D( {y_0})=1.
\end{eqnarray}
Hence, in order to obtain the phase space vector $\mathbf{z}$ at a given $ {y}$ value, all one should do is to solve the equation set (\ref{eq:evol_ABCD}) for the unknowns $\{A, B, C, D\}$ and substitute into (\ref{eq:T}) by considering (\ref{eq:T_block}).
\subsection{Observables}\label{Observables}
Now we want to demonstrate the physical relevance of the $\{A, B, C, D\}$ scalars.  {In this section, we will adopt Torre's viewpoint, in which two types of rays are identified by their initial conditions \citep{Torre:2005}}. Axial rays are  {defined as} those with $\{q_{in}=0, \theta _{in}=1/n_0\}$ and field rays have $\{q_{in}=1, \theta _{in}=0\}$. Therefore, $B$ and $D$ are representatives of  {the evolution of} axial rays; whereas, $A$ and $C$ represent the  {the evolution of} field rays. Thus, at any point of evolution, the $ABCD$ matrix represents a ray  {transformation} which is a superposition of  {the propagation of} an axial and a field ray. 

For the design of an optical system, one is usually interested in the magnification provided by the system, its primary and secondary focal lengths, power of the system etc. Here we will point some of those properties that will be relevant for our investigation in the relativistic case. For instance, there are two types of magnifications associated with an optical system: 1) ray-coordinate magnification, $M_q=q/q_{in}$ and ii) momentum magnification, $M_p=p/p_{in}$. In terms of the elements of the ray transfer matrix, they are given by \citep{Torre:2005}
\begin{eqnarray}
M_q=A+B\frac{p_{in}}{q_{in}},\qquad M_p=D+C\frac{q_{in}}{p_{in}}.
\end{eqnarray}  
For an axial ray, for example, $M_p$ is solely determined by the scalar $D$ and for a field ray $M_q$ is determined by $A$ only.

Let us say we have an optical system in between two mediums with different refractive indices $\none$ and $\ntwo$. Primary and secondary focal lengths are defined with respect to primary and secondary principal points respectively for the axial rays. The primary focal length is given by
\begin{eqnarray}\label{eq:fone_def}
{\fone}= \frac{q_{out}}{\tan \theta}\approx \frac{q_{out}}{\theta}=\frac{q_{out}}{p_{in}/\none_0},
\end{eqnarray} 
where $q_{out}$ and $p_{in}$ are the position and momentum variables at the output and input planes respectively.
In order to obtain $q_{out}$ consider the symplectic transfer matrix 
\begin{eqnarray}\label{eq:Transferfocals}
\left[
\begin{array}{c}
q_{out} \\
p_{out}
\end{array}
\right]&=&
\left[
\begin{array}{c c}
A({ {y_{out}}},{ {y_{in}}}) & B({ {y_{out}}},{ {y_{in}}}) \\
C({ {y_{out}}},{ {y_{in}}}) & D({ {y_{out}}},{ {y_{in}}})
\end{array}
\right]\,
\left[
\begin{array}{c}
q_{in}=0 \\
p_{in}
\end{array}
\right]\nonumber\\
\end{eqnarray}
Then through (\ref{eq:fone_def})
\begin{eqnarray}\label{eq:fone}
{\fone}=B({ {y_{out}}},{ {y_{in}}}){\none_0}.
\end{eqnarray}
Likewise, the magnitude of the secondary focal length is given by
\begin{eqnarray}\label{eq:ftwo}
{\ftwo}=- \frac{q_{in}}{p_{out}/\ntwo _0} =-{\ntwo_0}\left[ -B({ {y_{out}}},{ {y_{in}}})\right],
\end{eqnarray}
which follows from taking the inverse of the transfer matrix given in (\ref{eq:Transferfocals}). Then, 
\begin{eqnarray}\label{eq:fone_ftwo_ratio}
\frac{\fone}{\ftwo}=\frac{\none_0}{\ntwo_0}.
\end{eqnarray}
If the two mediums are the same, then of course $\fone=\ftwo$ holds. We will refer to this result in Section~\ref{Distances, reciprocity and symplecticity} when we discuss Etherington's distance reciprocity in the relativistic case.
\section{First order phase space ray optics for curved background}\label{First order phase space ray optics for curved background}

\subsection{Main idea}
The following are the guidelines for our construction of reduced phase space optics.
\begin{itemize}
\item[(i)] Our aim is to construct a phase space analogous to the one of Newtonian optical phase space in the paraxial regime. The Newtonian limit of our construction holds at the first order approximation. This is the regime which is mostly relevant for the cosmological and astrophysical distance calculations.
\item[(ii)] We do not directly refer to a $3+1$ decomposition of the underlying spacetime geometry. This is the approach, for example, that is used in order to find an analogue refractive index for the gravitational lensing spacetime which recovers the Newtonian limit up to full order. However, as it is seen at the previous section, only the up to second order Taylor expansion of the refractive index is relevant for Newtonian ray transfer matrices. 
\item[(iii)] In relativity, physically meaningful quantities are obtained once a fiducial worldline is introduced in the problem. In fact, this is not different for Newtonian optics: the eikonal equation, (\ref{eq:eikonal}), which results from Fermat's principle, is nothing but the geodesic equation of the optical metric,  {$ds^2_{opt.}=n^2ds^2$}; and the reduced phase space coordinates are defined \textit{with respect to the optical axis}. The fact that the optical axis is indeed another solution of the geodesic equation of the optical metric is usually overlooked. Thus, we apply null geodesic actions for two neighboring ray trajectories one of which serves as an optical axis, though, not an absolute one in the relativistic case. 
\item[(iv)] The methodology we follow here is constructed on Vines' derivation of geodesic deviation equation for high orders \citep{Vines:2014}. According to his work, a neighboring curve can be covariantly defined by making use of a fiducial geodesic and its exponential map\footnote{Actually, this idea was previously triggered by Aleksandrov and Piragas \citep{Aleksandrov:1979}. Also Ba\.{z}a\'{n}ski \citep{Bazanski:1977} had a similar construction for nonnull curves.}. This is done by introducing geodesic deviation bivectors defined through Synge's world function \citep{Synge:1960}. We aim to construct a phase space relevant for observations. Moreover, physical sizes of the objects on the sky are estimated by the proper sizes. Therefore, the world function, being the measure of proper distance between two spacetime points, is the most relevant tool for our construction.
\item[(v)] We pick a tetrad approach so that the underlying equations of motion are written in terms of the observables themselves.
\item[(vi)] Vines' action, up to quadratic order, is used to define a tetrad screen action that generates the underlying Lagrangian formalism. This quadratic Lagrangian allows us to pick physically relevant phase space (Darboux) coordinates. Note that this is inherently different to other constructions in relativistic optics in which the spatial spacetime coordinates are chosen as phase space coordinates and the ray momentum itself is chosen as the phase space canonical momentum (cf.  \citep{Perlick:2000}). 
\item[(vii)] We switch to a Hamiltonian formalism, define a quadratic Hamiltonian function and its corresponding Lie operator that evolves the system. Lie operators that are constructed via quadratic polynomials have matrix representations. That  is how we obtain a symplectic ray bundle transfer matrix that takes an initial phase space vector to a final one.
\end{itemize}
The next subsection, \ref{Our_Fermat_Synge}, will essentially be a brief summary of Vines' work \citep{Vines:2014} applied to our investigation which does not intend to recover his results fully. 
\subsection{Null geodesic actions with Synge's world function}\label{Our_Fermat_Synge}
 {As an analogous construction to the Newtonian ray optics, our aim in this section is to develop an action principle for null bundles within general relativity. The action in question will be our starting point in the formulation of symplectic transfer matrices of thin ray bundles. For this purpose, now we  introduce Synge's world function.}

Synge's world function $\sigma (r,s)$ depends on two spacetime points $r$ and $s$ which are connected by a unique geodesic, $\Gamma$, such that \citep{Synge:1960}
\[
    \sigma (r,s)= \frac{1}{2}
\begin{cases}
    (\rm{proper\,distance})^2,& \Gamma:\rm{spacelike}\\
    0,              &  \Gamma:\rm{null}\\
    -(\rm{proper\,time})^2,& \Gamma:\rm{timelike}.
\end{cases}
\]

 {Our initial aim here is to define a curve $\Lambda$ via a fiducial geodesic $\Sigma$ and an exponential map that is defined on $\Gamma$.
In Synge's formalism, one introduces \textit{bitensors} which depend on two spacetime points as connecting $\Sigma$, to a curve $\Lambda$ via $\Gamma$ is essentially nonlocal.} In our investigation, $\Gamma$ is spacelike and the length of $\Gamma$ is not assumed to be infinitesimally small in general. Therefore, one uses different coordinate indices for different spacetime points. Namely, for tensors defined at point $r$ we will use indices $\{\alpha, \beta, \gamma, \delta\}$ and for the ones defined at point $s$ we will use $\{\mu, \nu, \rho\}$.

Let us define the tangent vectors of $\Gamma$ defined at points $r$ and $s$ respectively as
\begin{eqnarray}
t^\alpha=\frac{dx^\alpha}{d\lambda}\Bigg|_{\lambda=\lambda_r}\,\,{\rm{and}}\,\,\,\, t^\mu=\frac{dx^\mu}{d\lambda}\Bigg|_{\lambda=\lambda_s},
\end{eqnarray} 
where $\lambda$ is an affine parameter which puts the geodesic equation of $\Gamma$ into the $\nabla_{\vec{t}}\vec{t}=0$ form and subscripts assigned to $\lambda$ refers to its value at a given point. Our proper length is then
\begin{eqnarray}
\sigma=\frac{1}{2}\left(\Delta \lambda \right)^2t^2,
\end{eqnarray}
where $\Delta \lambda$ is not necessarily small and $t^2=t^\alpha t_\alpha=t^\mu t_\mu$ as the tangent vector is parallel transported on $\Gamma$.

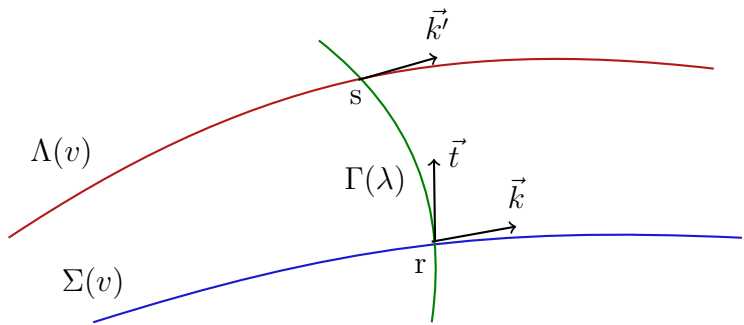
\begin{figure}
\hspace*{1cm}
\begin{center}
\begin{tikzpicture}[thick, scale=1.5]
\draw [myred] (1.25,1.5) to[bend left=20] (7.50,3);
\draw [myblue] (2,0.75) to[bend left=10] (7.75,1.5);
\draw [mygreen] (5,0.75) to[bend right] (4,3.25);

\node [draw=none] at (4.9,1.25) {r};
\node [draw=none] at (4.33,2.75) {s};

\node [draw=none] at (4.5,2) {$\Gamma(\lambda)$};
\node [draw=none] at (2,1.1) { $\Sigma (v)$};
\node [draw=none] at (1.72,2.25) {$\Lambda (v)$};

\draw [->] (5,1.465) to (5.75,1.6);
\draw [->] (5.03,1.465) to (5.02,2.2);
\draw [->] (4.35,2.9) to (5.05,3.1);

\node [draw=none] at (5.75,1.9) {$\vec{k}$};
\node [draw=none] at (5.05,3.4) {$\vec{k'}$};
\node [draw=none] at (5.2,2.25) {$\vec{t}$};
\end{tikzpicture}
\end{center}
\caption{The fiducial null geodesic $\Sigma (v)$ is plotted in blue. The green curve represents the spacelike $\Gamma(\lambda)$ given by Synge's world function. The null geodesic $\Lambda (v)$ is in red which can be uniquely obtained through $\Sigma (v)$ and $\Gamma(\lambda)$.}\label{fig:Syngesworldfun}
\end{figure}

Now, we will pick a fiducial geodesic, $\Sigma$, which can have any causal character in Vines' construction but will be null in our case. We will identify point $r$ as the intersection of $\Gamma(\lambda)$ and $\Sigma (v)$, in which $v$ is the affine parameter that puts the geodesic equation into $\nabla_{\vec{k}}\vec{k}=0$ form for the tangent vector, $\vec{k}$, of $\Sigma$ (See figure~\ref{fig:Syngesworldfun}.). Moreover, we will define another curve $\Lambda$, which can, again, have any causal character and does not even have to be a geodesic in Vines' work but will be a null geodesic in our case. We pick an isochronous correspondence such that $\Lambda=\Lambda (v)$ and the tangent vector, $\vec{k'}$, satisfies $\nabla_{\vec{k'}}\vec{k'}=0$. Similarly, we will identify the point $s$ as the intersection of $\Gamma(\lambda)$ and $\Lambda (v)$. 

 {Next, we define an exponential map
\begin{eqnarray}
\xi^\alpha(v)=-\sigma ^\alpha\left(r(v),s(v)\right),
\end{eqnarray}
on $\Gamma(v)$ and consider a null fiducial geodesic, $\Sigma(v)$, in order to specify another null geodesic $\Lambda(v)$.}
Note that, due to its nonlocal nature, $\xi^\alpha$ acts as a vector with respect to the tensorial operations conducted at $x^\beta$, however, it acts as a scalar with respect to those operations conducted at $x^\mu$ \citep{Poisson:2003}. The term $\sigma ^\alpha=\nabla^\alpha \sigma=-\Delta \lambda \,t^\alpha$ is simply the covariant derivative of the world function at point $r$. Similarly, $\sigma ^\mu=\nabla^\mu \sigma=\Delta \lambda \,t^\mu$ is its covariant derivative at point $s$, such that
\begin{eqnarray}\label{eq:sigma_magnitude}
\sigma^\mu\sigma_\mu=2\sigma=\sigma^\alpha\sigma_\alpha.
\end{eqnarray}

We would like to know how $\xi^\alpha$ changes with respect to the parameter $v$. The nonlocal nature appears in the definition of the total covariant $v-$derivative as well. The derivation is taken with respect to the spacetime covariant derivatives defined \textit{both} at $x^\beta$ and at $x^\mu$. It is given by
\begin{eqnarray}\label{eq:Total_cov_der_xi}
\dot{\xi}^\alpha=\frac{\mathbb{D}\xi^\alpha}{dv}=-\left(k^\beta\nabla_\beta+{k'}^\mu\nabla_\mu\right)\sigma ^\alpha.
\end{eqnarray} 
Moreover, we want to write (\ref{eq:Total_cov_der_xi}) in terms of a given set $\{\vec{\xi}, \vec{k},\vec{{k'}}\}$. For this, Vines considers the following analogy.
In flat space, an ordinary function defined at a point can be written in terms of the powers of the coordinate displacement vector via an ordinary Taylor expansion. Similarly, one can covariantly expand (\ref{eq:Total_cov_der_xi}) in powers of $\sigma^\alpha (r,s)$ at the coincidence limit $r\rightarrow s$ as it acts like a nonlocal displacement vector in general. Then, one writes the expanded $\dot{\xi}^\alpha$ as \citep{Vines:2014}
\begin{eqnarray}\label{eq:Dot_xi_expanded}
\dot{\xi}^\alpha =-k^\beta\left(\tensor{\delta}{^\alpha _\beta}-\frac{1}{3}\tensor{R}{^\alpha _{\vec{\xi}}_\beta_{\vec{\xi}}}\right)
+{k'}^\mu \tensor{g}{^\beta _{\mu}}\left(\tensor{\delta}{^\alpha _\beta}+\frac{1}{6}\tensor{R}{^\alpha _{\vec{\xi}}_\beta_{\vec{\xi}}}\right)+O(\vec{\xi}^3).
\end{eqnarray}
Here $\tensor{g}{^\beta _{\mu}}$ is the \textit{parallel propagator}\footnote{
Parallel transport, $V^\mu$, of an arbitrary vector $V^\alpha$ defined at point $r$ along $\Gamma$ is given by 
\begin{equation*}
V^\mu=\tensor{g}{^\mu_\alpha}\left(s,r\right)V^\alpha.
\end{equation*}
Here $\tensor{g}{^\mu_\alpha}$ is defined by \citep{Poisson:2003}
\begin{equation*}
\tensor{g}{^\mu_\alpha}\left(s,r\right)=e_{{A}}^\mu(s)e^{{A}}_\alpha(r),
\end{equation*}
in which $e_{{A}}^\mu(s)$ and $e_{{A}}^\alpha(r)$ are the local orthonormal tetrad fields defined at points $s$ and $r$ respectively, such that 
\begin{equation*}
\tensor{g}{_\mu_\nu}e_{{A}}^\mu e_{{B}}^\nu=\tensor{\eta}{_{{A}}_{{B}}}=\tensor{g}{_\alpha_\beta}e_{{A}}^\alpha e_{{B}}^\beta,
\end{equation*}
with $\tensor{\eta}{_{{A}}_{{B}}}=\rm{diag}\left(-1,1,1,1\right)$.}
and $\tensor{R}{^\alpha _{\vec{\xi}}_\beta_{\vec{\xi}}}=\tensor{R}{^\alpha _{\gamma}_\beta_{\delta}}\xi ^\gamma \xi^\delta$ with $\tensor{R}{^\alpha _{\gamma}_\beta_{\delta}}$ being the Riemann curvature tensor. The terms in the parenthesis follow from the second variation of the world function conducted at points $r$ and $s$, respectively. Now, as claimed before, one can write ${k'}^\mu$ in terms of the deviation vector variables and the tangent vector of a fiducial null geodesic, $k^\alpha$. This is obtained by solving (\ref{eq:Dot_xi_expanded}) via a perturbative approach in $\xi ^\alpha$ for ${k'^\mu}$, i.e.,
\begin{eqnarray}\label{eq:k_prime}
{k'}^\mu=\tensor{g}{^\mu_\alpha}\left(k^\alpha+\dot{\xi}^\alpha-\frac{1}{2}\tensor{R}{^\alpha _{\vec{\xi}}_{\vec{k}}_{\vec{\xi}}}\right)+\dot{\vec{\xi}}\cdot O(\vec{\xi}^2)+O(\vec{\xi}^3).
\end{eqnarray}

Let us now write the null geodesic action \citep{Schneider:1999} for the curve $\Lambda$ by using (\ref{eq:k_prime})
\begin{eqnarray}\label{eq:S_Lambda}
S_{\Lambda}&=&\int \frac{1}{2}{k'}^2 dv \\ \nonumber
 &=& S_{\Sigma} + \int \frac{1}{2}\left[2\vec{k}\cdot \dot{\vec{\xi}}+\dot{\vec{\xi}}\cdot \dot{\vec{\xi}}-\tensor{R}{_{\vec{\xi}}_{\vec{k}}_{\vec{\xi}}_{\vec{k}}}+O(\vec{\xi}, \dot{\vec{\xi}})^3\right]dv,
\end{eqnarray}
where $S_{\Sigma}=\int \frac{1}{2} k^2 dv$ is the geodesic action for the null curve $\Sigma$. 

As mentioned in the previous subsection, in relativity, physically meaningful quantities are obtained once the fiducial motion is introduced into the problem. Therefore, we claim that null geodesic action should be applied more than once, as we do here for $S_{\Lambda}$ and $S_{\Sigma}$, in order to get a well defined action principle for physically relevant optical quantities. In the relativistic case, it is the integral curves of the null vector $\vec{k}$, i.e., the central geodesic $\Sigma$, that plays the role of the optical axis. The neighboring null vector $\vec{k'}$ can then be interpreted as the tangent vector of the outermost ray of a null congruence. 

Recall that our aim is to define the first order ray propagation in relativity with transfer matrices analogous to the case in the Newtonian paraxial regime. Therefore, from now on, we will assume that $\Lambda$ and $\Sigma$ are nearly parallel neighboring null geodesics such that $\Delta \lambda$ is small. Accordingly, we will keep the terms up to quadratic order in the action (\ref{eq:S_Lambda}). As $\delta S_\Sigma=0$ and the $\vec{k}\cdot \dot{\vec{\xi}}$ term in $S_\Lambda$ is a total derivative, Vines chooses to omit these terms in the action. In the next section, we will show that $\vec{k}\cdot \dot{\vec{\xi}}$ term is indeed zero for our physical problem. Therefore, we write the geodesic action for $\vec{k'}$ with respect to the neighboring null geodesic and up to quadratic order as
\begin{eqnarray}\label{eq:The_action}
S=\int \left(\frac{1}{2}\dot{\vec{\xi}}\cdot \dot{\vec{\xi}}+\frac{1}{2}\tensor{R}{_{\vec{\xi}}_{\vec{k}}_{\vec{k}}_{\vec{\xi}}}\right)dv,
\end{eqnarray}
where we have we made use of Riemann tensor symmetries on the second term and omit the subscript $\Lambda$ in the notation for convenience. {Note that for the case of bilocal objects in general, one considers taking successive symmetrized covariant derivatives in order to obtain the local, observable characteristic of a tensor field \citep{Aleksandrov:1979}. We follow this common practice here and the} overdot now represents the covariant derivative with respect to $\vec{k}$, i.e., $\dot{\vec{\xi}}=D\vec{\xi}/dv=\nabla _{\vec{k}}\vec{\xi}$ due to our small deviation assumption.

Following (\ref{eq:The_action}) we will take 
\begin{eqnarray}\label{eq:Lagrangian_Vines}
\tilde{\mathcal{L}}=\frac{1}{2}\dot{\vec{\xi}}\cdot \dot{\vec{\xi}}+\frac{1}{2}\tensor{R}{_{\vec{\xi}}_{\vec{k}}_{\vec{k}}_{\vec{\xi}}}
\end{eqnarray}
as our Lagrangian function and $v$ as our evolution parameter. Then varying (\ref{eq:The_action}) with respect to both  {$\vec{\xi} $ and $\dot{\vec{\xi}}$} yields the equations of motion
\begin{eqnarray}\label{eq:first_ord_dev}
\ddot{\xi}^\alpha=\tensor{R}{^\alpha_{\vec{k}}_{\vec{k}}_{\vec{\xi}}}\,,
\end{eqnarray}
which is just the first order approximation of the geodesic deviation equation that is often mistakenly referred to as \textit{the} geodesic deviation equation in the literature.
 {\subsection{Reduced Lagrangian and the screen basis}\label{Reduced Lagrangian and the screen basis}}
 {Recall from Section~\ref{First order Newtonian ray optics} that in order to obtain the ray transfer matrix of first order Newtonian optics in the paraxial regime, one reduces the system by one order.  In this subsection,  we will demonstrate how the Lagrangian (\ref{eq:Lagrangian_Vines}) can naturally be reduced in the general relativistic setting. This will allow us to transform to a Hamiltonian formulation in a reduced phase space in which one can represent an observed thin ray bundle evolution by a first order symplectic transformation.}

 {Assume that the central ray of an observed null bundle is given by $k^\alpha = \omega \left(u^\alpha  + r^\alpha \right)$. Here $\omega=-\vec{k} \cdot \vec{u}$ is the value of the frequency of light measured by an observer with 4-velocity $u^\alpha$. Vector $r^\alpha$ is along the spatial direction of the null ray that satisfies $\vec{u}\cdot \vec{r}=0$. Such a decomposition of $\vec{k}$ is typical within studies that involve the investigation of observed null bundles \citep{Ellis:2012}. We will also set the value of the frequency as $\omega _o=1$ at the measurement point as it is done in various applications in the literature.}

 {Note that in the first order limit of the geodesic deviation, the geodesic vector $\vec{k}$ and the corresponding Jacobi field $\vec{\xi}$ are assumed to be Lie dragged along the integral curves of each other, i.e., $\nabla _{\vec{k}}\vec{\xi}=\nabla _{\vec{\xi}}\vec{k}$. This condition, first order geodesic deviation equation and the fact that $\vec{k}$ is null, guarantee that ${\vec{k}}\cdot\vec{\xi}={\rm{constant}}$. Since we are interested in a physical problem in which there is always an observation point on the bundle, $\vec{k}\cdot\vec{\xi}=0$ holds initially and, given the argument above, throughout the propagation of the bundle. Then, our deviation vector swipes the null cone throughout the evolution \citep{Perlick:2010} and it can be decomposed into components {with respect to the degrees of freedom of the thin bundle as}
\begin{eqnarray}
\vec{\xi}=\xi^{k}\vec{k}+\boldsymbol{\xi},
\end{eqnarray}
in which
\begin{eqnarray}
\boldsymbol{\xi}=\xi^{{1}}\vec{s}_{{1}}+\xi^{{2}}\vec{s}_{{2}}.
\end{eqnarray}
Those are the $s^\alpha_{a}$ basis components of the deviation vector that live on our observational screen with $\{a,b\}=\{1,2\}$. This dyad is assumed to be $C^\infty$ along the null geodesic. For our construction, we pick such a basis that 
\begin{eqnarray}
\vec{s}_{a}\cdot \vec{s}_{b}=\delta _{a b},
\qquad
\vec{u}\cdot\vec{s}_{a}=0,
\qquad
\vec{r}\cdot\vec{s}_{a}=0,
\qquad
\nabla _{\vec{k}}\vec{s}_{a}=0, \nonumber
\end{eqnarray}
are satisfied. Then $s^\alpha_{a}$ forms the \textit{Sachs basis} \citep{Sachs:1961} that is parallel propagated along the central light ray. This guarantees that the 2-dimensional spatial screen, on which the observables are projected, refers to the same screen at each point of the light propagation. Sachs basis is defined uniquely up to rotations around the spatial vector $\vec{r}$. That means, for a constant orthogonal matrix, $\tensor{\mathcal{O}}{^b_a}$, the basis transforms as $\vec{s}_a=\tensor{\mathcal{O}}{^b_a}\vec{s}_b$ with $\tensor{\mathcal{O}}{^b_a}\tensor{\mathcal{O}}{^d_c}\delta _{bd}=\delta _{ac}$ \citep{Perlick:2000}.}

{We should also mention that under such a construction, $\nabla _{\vec{k}}\vec{u}=0$ needs to be satisfied in order for the Sachs basis to remain orthogonal to the null ray throughout the evolution. Note that we are free to pick such set of observers as the bundle morphology is independent of the choice of the observers\citep{Schneider:1999}.}\footnote{ {For less strict conditions on an observational screen see for example \citep{Fanizza:2013}.}} 

 {Then, given such a Sachs basis, it is easy to show that the $\vec{k}\cdot \dot{\vec{\xi}}$ term that appears in the geodesic action (\ref{eq:S_Lambda}) becomes $\vec{k}\cdot \dot{\vec{\xi}}=\vec{k}\cdot \boldsymbol{\dot{\xi}}=0$. Moreover, the first term that appears in the Lagrangian (\ref{eq:Lagrangian_Vines}) can be written as
\begin{eqnarray}
\dot{\vec{\xi}}\cdot \dot{\vec{\xi}}=\tensor{\delta}{_{a}_{b}} \dot{\xi}^{a}\dot{\xi}^{b}.
\end{eqnarray}
Likewise, the second term in (\ref{eq:Lagrangian_Vines}) follows as
\begin{eqnarray}
\tensor{R}{_{\vec{\xi}}_{\vec{k}}_{\vec{k}}_{\vec{\xi}}}=\tensor{R}{_{a}_{\vec{k}}_{\vec{k}}_{b}}\xi^{a}\xi^{b}\equiv \tensor{R}{_{\boldsymbol{\xi}}_{\vec{k}}_{\vec{k}}_{\boldsymbol{\xi}}},
\end{eqnarray}
due to $\tensor{R}{_{\vec{k}}_{\vec{k}}_{\vec{k}}_{\vec{k}}}$ and $\tensor{R} {_{\vec{k}}_{\vec{k}}_{\vec{k}}_{a}}$ being zero.} Then, only two degrees of freedom survive in $\tilde{\mathcal{L}}$ and we write the \textit{reduced Lagrangian} as
\begin{eqnarray}\label{eq:Red_Lagrangian}
L=\frac{1}{2}\tensor{\delta}{_{a}_{b}} \dot{\xi}^{a}\dot{\xi}^{b}+\frac{1}{2}\tensor{R}{_{a}_{\vec{k}}_{\vec{k}}_{b}}\xi^{a}\xi^{b}.
\end{eqnarray}
The term $\tensor{\mathcal{R}}{_{\,}_{a}_{b}}:=\tensor{R}{_{a}_{\vec{k}}_{\vec{k}}_{b}}$ is usually referred to as \textit{optical tidal matrix} in cosmological light propagation studies \citep{Seitz:1994, Perlick:2010}. The overdot that appears in (\ref{eq:Red_Lagrangian}) now denotes a simple total derivative with respect to the affine parameter $v$ as we consider the \textit{ {dyad} components} of the deviation vector here.

\subsection{Reduced Hamiltonian and ABCD Matrices}\label{Reduced Hamiltonian and ABCD Matrices}
Let us define a 4-dimensional symplectic phase space $M(\mathbb{R}^{4})$. We will denote the phase space coordinates and the momenta canonically conjugate to them that follow from the reduced Lagrangian (\ref{eq:Red_Lagrangian}) as
\begin{eqnarray}
q^{a}&=&\xi ^{a}\\ \nonumber
p_{a}&=&\frac{\partial L}{\partial \dot{q}^{a}}=\dot{\xi} _{a}.
\end{eqnarray}
Then we can define a reduced Hamiltonian function via
\begin{eqnarray}\label{eq:Red_Hamiltonian}
H=p_{a}\dot{q}^{a}-L
=\frac{1}{2}\tensor{\delta}{^{a}^{b}} \dot{\xi}_{a}\dot{\xi}_{b}-\frac{1}{2}\tensor{\mathcal{R}}{_{\,}_{a}_{b}}\xi^{a}\xi^{b}.
\end{eqnarray}
Note that the reduced Hamiltonian (\ref{eq:Red_Hamiltonian}) is analogous to the Newtonian one given in (\ref{eq:Hquad}) with $\tensor{\mathcal{R}}{_{\,}_{a}_{b}}$ being analogous to $\tensor{n}{_a_b}$, i.e., second variation of the refractive index. This is no surprise as light propagation within a medium of refractive index $n$ with Euclidean metric components $\delta _{\mu \nu}$, in fact corresponds to a propagation through a curved background with the optical metric components $g_{\mu \nu}=n^2\delta _{\mu \nu}$. Then, second variation of $g_{\mu \nu}$ are given by the Riemann tensor components.

We would also like to emphasize that in the Newtonian case that we presented in Section~\ref{First order Newtonian ray optics}, the propagation vector $\vec{k}$ is spacelike and $\tensor{R}{_{\boldsymbol{\xi}}_{\vec{k}}_{\vec{k}}_{\boldsymbol{\xi}}}$ indeed represents the Gaussian curvature, $K_0$, of a 2-dimensional subspace defined by $\vec{k}$ and $\boldsymbol{\xi}$ - up to the squared area of the corresponding parallelogram, i.e.,
\begin{eqnarray}
K_0=\frac{-\tensor{R}{_{\boldsymbol{\xi}}_{\vec{k}}_{\vec{k}}_{\boldsymbol{\xi}}}}{\Bigl[g\left(\boldsymbol{\xi},\boldsymbol{\xi}\right)g\left(\vec{k},\vec{k}\right)-g\left(\boldsymbol{\xi},\vec{k}\right)g\left(\vec{k},\boldsymbol{\xi}\right)\Bigr]}.
\end{eqnarray}
This explains why $\tensor{n}{_a_b}$ (or $n_2$) term has such fundamental importance in the GRIN profiles for light propagation or fiber-optics studies as we discussed in footnote (\ref{fn:GRIN}). For the case of general relativity, $\vec{k}$ is null and the corresponding 2-dimensional subspace is referred to as the \textit{half light-like surface} \citep{Duggal:2011}. In that case, $\tensor{R}{_{\boldsymbol{\xi}}_{\vec{k}}_{\vec{k}}_{\boldsymbol{\xi}}}$ is a measure of null sectional curvature, $K_{\vec{k}}$, that is given by \citep{Beem:1996}
\begin{eqnarray}
K_{\vec{k}}=\frac{-\tensor{R}{_{\boldsymbol{\xi}}_{\vec{k}}_{\vec{k}}_{\boldsymbol{\xi}}}}{g\left(\boldsymbol{\xi},\boldsymbol{\xi}\right)}.
\end{eqnarray}

Let us now return to our original problem and write the Hamilton-Jacobi equations in the following form
\begin{eqnarray}\label{eq:Hamilton_Jacobi_z}
\frac{dz^i}{dv}=\Omega ^{ij}\frac{\partial {H}}{\partial z^j}=-\{{H},z^i\},
\end{eqnarray}
in which the phase space vector components are
\begin{eqnarray}\label{eq:Phase_coords}
\mathbf{z}=
\left[\begin{array}{c }
 q ^{a} \\
 p _{b} 
\end{array}\right]
=\left[\begin{array}{c }
 \xi ^{{1}} \\
 \xi ^{{2}} \\
 \dot{\xi} _{{1}} \\
 \dot{\xi} _{{2}}
\end{array}\right],
\end{eqnarray}
and $\mathbf{\Omega}$ is the fundamental symplectic matrix defined in (\ref{eq:Poisson_z})-(\ref{eq:Omega_prop}) before.

Now we will define a Lie operator associated with the reduced Hamiltonian (\ref{eq:Red_Hamiltonian}) as
\begin{eqnarray}\label{eq:Lie_op}
\hat{\mathcal{L}}_{{H}}\left[\bullet \right]&=&-\{{H},\bullet \}=
-\frac{\partial {H}}{\partial z^i}\Omega ^{ij}\frac{\partial}{\partial z^j}\\ \nonumber
&=&\delta ^{ab}\dot{\xi}_b\frac{\partial}{\partial \xi^a}+\tensor{\mathcal{R}}{_{\,}_{a}_{b}}\xi^b\frac{\partial}{\partial \dot{\xi}_a},
\end{eqnarray}
which is analogous to the Lie operator of an attractive or a repulsive harmonic oscillator depending on the sign of $\tensor{\mathcal{R}}{_{\,}_{a}_{b}}$.

Note that our Hamiltonian vector field ${H}^i=\Omega ^{ij}\partial {H}/\partial z^j$ is curl-free, i.e., $\partial ^i{H}^j-\partial^j{H}^i=0$ and it represents a linear Hamiltonian flow. This is possible due to: (i) ${H}$ being written up to quadratic order with respect to phase space coordinates, (ii) the Riemann tensor having certain symmetries, namely $\tensor{\mathcal{R}}{_{\,}_{a}_{b}}=\tensor{\mathcal{R}}{_{\,}_{b}_{a}}$. Therefore, we can define a $4\times 4$ Hamiltonian matrix, $\mathbf{L}_{\mathbf{{H}}}$, which is the representation of the Lie operator (\ref{eq:Lie_op}) that we write as
\begin{eqnarray}\label{eq:Lie_matrix}
\mathbf{L}_{\mathbf{{H}}}=
\left[
\begin{array}{c|c}
\mathbf{0_2} & \, \, \delta ^{ab} \\
\hline
\tensor{\mathcal{R}}{_{\,}_{a}_{b}} & \, \, \mathbf{0_2}
\end{array}
\right].
\end{eqnarray}
Our Lie operator (\ref{eq:Lie_op}) and its matrix representation (\ref{eq:Lie_matrix}) applicable for a curved background are analogous to (\ref{eq:Lie_op_Newtonian}) and (\ref{eq:Lie_matrix_Newtonian}) given in the Newtonian case.

Next, we rewrite the Hamilton-Jacobi equations (\ref{eq:Hamilton_Jacobi_z}) in the matrix form
\begin{eqnarray}\label{eq:Ham_eqs_matrix_Gr}
\dot{\mathbf{z}}=\mathbf{L}_{\mathbf{{H}}}\,\mathbf{z}.
\end{eqnarray}
The matrix $\mathbf{L}_{\mathbf{{H}}}$ is the generator of the infinitesimal evolution, i.e.,
\begin{eqnarray}
\mathbf{z}(v+dv)=\exp{[{\mathbf{L}_{\mathbf{{H}}}dv}]}\mathbf{z}(v).
\end{eqnarray}
The evolution of the system between any initial and final points is then obtained by the linear transformation of the phase space vector, i.e.,
\begin{eqnarray}\label{eq:TransferGr}
\mathbf{z}=\mathbf{T}\left(v,v_0\right)\mathbf{z_0},
\end{eqnarray}
in which $\mathbf{T}$ is the \textit{ray bundle transfer matrix}. As in Section~\ref{Symplectic geometry and ABCD matrices}, it is determined by substituting (\ref{eq:TransferGr}) into (\ref{eq:Ham_eqs_matrix_Gr}) so that we have 
\begin{eqnarray}\label{eq:Hamilton_T_matrix}
\dot{\mathbf{T}}=\mathbf{L}_{\mathbf{{H}}}\,\mathbf{T}.
\end{eqnarray}
Its solution is
\begin{eqnarray}\label{eq:Transfer_OE}
\mathbf{T}\left(v,v_0\right)={\rm{OE}}\left[{\int _{v_0}^{v}\mathbf{L}_{\mathbf{{H}}}dv}\right]\mathbf{T}\left(v_0,v_0\right),
\end{eqnarray}
with initial conditions $\mathbf{T}\left(v_0,v_0\right)=\mathbf{I_4}$.
Note that the optical tidal matrix, $\boldsymbol{\mathcal{R}}$, is $v$-dependent for a generic spacetime and the corresponding Lie operators do not commute at different points unless the underlying spacetime has some nice symmetry properties. Therefore, determination of $\mathbf{T}$ involves an ordered exponentiation (OE) with respect to the affine parameter $v$. 

We will write the ray bundle transfer matrix in an $ABCD$ block form
\begin{eqnarray}\label{eq:Transfer_block_GR}
\mathbf{T}=
\left[
\begin{array}{c|c}
\mathbf{A} & \mathbf{B} \\
\hline
\mathbf{C} & \mathbf{D}
\end{array}
\right].
\end{eqnarray}
Note that $\mathbf{T}$ is a symplectic matrix which satisfies (\ref{eq:Symplectic_T}). Then substitution of (\ref{eq:Transfer_block_GR}) into (\ref{eq:Hamilton_T_matrix}) gives us a set of 16 equations
\begin{eqnarray}\label{eq:four_sets}
\mathbf{\dot{A}}&=&\mathbf{C},\qquad \,\,\,\, \mathbf{A}\left(v_0,v_0\right)=\mathbf{I_2},\\ \nonumber
\mathbf{\dot{B}}&=&\mathbf{D},\qquad \,\,\,\, \mathbf{B}\left(v_0,v_0\right)=\mathbf{0_2},\\ \nonumber
\mathbf{\dot{C}}&=&\boldsymbol{\mathcal{R}}\mathbf{A},\qquad  \mathbf{C}\left(v_0,v_0\right)=\mathbf{0_2},\\ \nonumber
\mathbf{\dot{D}}&=&\boldsymbol{\mathcal{R}}\mathbf{B},\qquad  \mathbf{D}\left(v_0,v_0\right)=\mathbf{I_2},\\ \nonumber
\end{eqnarray}
to solve in order to construct the ray bundle transfer matrix.
\section{Distances, reciprocity and symplecticity}\label{Distances, reciprocity and symplecticity}
 {\subsection{Cosmological distances and reciprocity}}
We will now link our construction with certain definitions and methods that already exist in the literature. Recall that in Section~\ref{Observables}, we  {mentioned} two types of rays: axial rays and field rays  {\citep{Torre:2005}}. For standard cosmological calculations, for example, one is usually interested in the solutions for axial rays such that the observation point is a vertex. In that case, one usually determines the angular diameter distance, ${D}_A$, and the luminosity distance, ${D}_L$, between the source and the observer which are respectively given by
\begin{eqnarray}\label{eq:defDADL}
{D}_A= \left(\frac{dS_s}{d\Theta _o}\right)^{1/2}\,\,\,\, {\rm{and}}\,\,\,\, {D}_L=\left( \frac{dS_o}{d\Theta _s}\right)^{1/2}.
\end{eqnarray}
Here $dS$ is the cross sectional area of the ray bundle evaluated at the source, $s$, or at the observation point, $o$, and likewise $d\Theta$'s are the solid angles. Those are obtained by \citep{Ellis:1971}
\begin{eqnarray}\label{eq:Areas_angles}
dS_s&:=&\left|{\xi} ^{{1}}\wedge {\xi} ^{{2}}\right|_s, \qquad dS_o:=\left|{\tilde{\xi}} ^{{1}}\wedge {\tilde{\xi}} ^{{2}}\right| _o,\\ \nonumber
d\Theta _o&:=&\left|\frac{d{\xi} ^{{1}}}{d\ell}\wedge \frac{d{\xi} ^{{2}}}{d\ell}\right|_o,\qquad
d\Theta _s:=\left|\frac{d{\tilde{\xi}} ^{{1}}}{d\ell}\wedge \frac{d{\tilde{\xi}} ^{{2}}}{d\ell}\right|_s,
\label{eq:dTheta}
\end{eqnarray}
in which $\wedge$ denotes the exterior product and $d\ell$ is the proper length. The Jacobi fields $\vec{\xi}$ and $\vec{\tilde{\xi}}$ correspond respectively to the bundles that are sent from point $o$ to $s$ and $s$ to $o$. We assume that these two bundles share the same central null geodesic. Note that the relation between $d\ell$ and the proper time $d\tau$ to the affine parameter $v$ is given by
\begin{eqnarray}
\left|d\ell \right|=\left|d\tau \right|=\left(-k^au_a\right)dv=\omega dv.
\end{eqnarray}
Now considering points $o$ and $s$ to be the respective measurement points, our ray bundle transfers follow as
\begin{eqnarray}\label{eq:Transfer_o_to_s}
\left[\begin{array}{c}
 \boldsymbol{\xi} \\
 \boldsymbol{\dot{\xi}}
\end{array}\right]_{s}= 
\left[
\begin{array}{c|c}
\mathbf{A} & \, \, \mathbf{B} \\
\hline
\mathbf{C} & \, \, \mathbf{D}
\end{array}
\right]_{(v_s,v_o)}
\left[\begin{array}{c}
 \boldsymbol{0} \\
 \boldsymbol{\dot{\xi}}
\end{array}\right]_{o},\,
\end{eqnarray}
and
\begin{eqnarray}\label{eq:Transfer_s_to_o}
\left[\begin{array}{c}
 \boldsymbol{\tilde{\xi}} \\
 \boldsymbol{\dot{\tilde{\xi}}}
\end{array}\right]_{o}= 
\left[
\begin{array}{c|c}
\mathbf{A} & \, \, \mathbf{B} \\
\hline
\mathbf{C} & \, \, \mathbf{D}
\end{array}
\right]_{(v_o,v_s)}
\left[\begin{array}{c}
 \boldsymbol{0} \\
 \boldsymbol{\dot{\tilde{\xi}}}
\end{array}\right]_{s},\,
\end{eqnarray}
Then following (\ref{eq:defDADL})-(\ref{eq:Areas_angles}) and (\ref{eq:Transfer_o_to_s})-(\ref{eq:Transfer_s_to_o}), one writes
\begin{eqnarray}\label{eq:D_A_D_L_B}
{D}_A=\omega _o{\rm{det}}\left|\mathbf{B}\left(v_s,v_o\right)\right|^{1/2},\qquad \rm{o-fixed}\\ \nonumber
{D}_L=\omega _s{\rm{det}}\left|\mathbf{B}\left(v_o,v_s\right)\right|^{1/2},\qquad \rm{s-fixed}
\end{eqnarray}
such that Etherington's distance reciprocity \citep{Etherington:1933}
\begin{eqnarray}\label{eq:Etherington}
D_L=\left(1+\mathpzc{z}\right)D_A
\end{eqnarray}
is satisfied with $\mathpzc{z}=\omega _s/\omega _o-1$ being the redshift. Note that matrix $\mathbf{B}$ is referred to as the \textit{Jacobi matrix} and it is usually denoted as $\mathcal{D}$ or $J$ in the literature. This is a good enough naming for light propagation with initial point being a vertex. However, for light propagation between any two points along the null path, it is the symplectic matrix $\mathbf{T}$ which is indeed the full Jacobi matrix. We observe that (\ref{eq:D_A_D_L_B}) and (\ref{eq:Etherington}) are analogous to (\ref{eq:fone})-(\ref{eq:fone_ftwo_ratio}) such that angular diameter and luminosity distances are analogous to the primary and secondary focal lengths of an optical system in the paraxial regime given in Section~\ref{Observables}. 

Note that, in the literature, one way of proving that (\ref{eq:Etherington}) follows from (\ref{eq:D_A_D_L_B}) is shown by \citep{Perlick:2010, Fleury:2015rwa} 
\begin{eqnarray}\label{eq:Recip_proof}
\mathbf{\dot{B}}\left(v,v_{ {o}}\right)\mathbf{{B}}^{\intercal}\left(v,v_{ {s}}\right)-\mathbf{{B}}^{\intercal}\left(v,v_{ {o}}\right)\mathbf{\dot{B}}\left(v,v_{ {s}}\right)
\end{eqnarray}
being a constant along the ray such that
\begin{eqnarray}\label{eq:Recip_B}
\mathbf{B}\left(v_{ {s}},v_{ {o}}\right)=-\mathbf{B}^{\intercal}\left(v_{ {o}},v_{ {s}}\right)
\end{eqnarray}
holds and the determinants in (\ref{eq:D_A_D_L_B}) have the same value. 

\begin{figure}
\hspace*{2.75cm}
\begin{tikzpicture}[thick,baseline=-0.5cm, scale=1.25]
\draw [myred] (9,0) to[bend right] (6,1.25);
\draw [myred] (6,0) to[bend right] (6,1.25);
\draw [myred] (6,0) to[bend left] (6,1.25);

\draw [myblue] (6,1.25) to[bend left] (3,1.75);
\draw [myblue] (3,0) to[bend right] (3,1.75);
\draw [myblue] (3,0) to[bend left] (3,1.75);

\draw [mygreen] (3,1.75) to[bend right] (0,2.5);
\draw [mygreen] (0,0) to[bend right] (0,2.5);
\draw [mygreen] (0,0) to[bend left] (0,2.5);

\draw [mygreen, dashed] (0,0) to[bend left] (3,1.1);
\draw [mygreen, dashed] (3,0) to[bend right] (3,1.1);
\draw [mygreen, dashed] (3,0) to[bend left] (3,1.1);

\draw [myblue, dashed] (3,1.1) to[bend right] (6,2);
\draw [myblue, dashed] (6,0) to[bend right] (6,2);
\draw [myblue, dashed] (6,0) to[bend left] (6,2);

\draw [myred, dashed] (6,2) to[bend left] (9,3);
\draw [myred, dashed] (9,0) to[bend right] (9,3);
\draw [myred, dashed] (9,0) to[bend left] (9,3);

\draw (0,0) -- (9,0);
\node [draw=none] at (0,-0.25) {s};
\node [draw=none] at (3,-0.25) {g};
\node [draw=none] at (6,-0.25) {h};
\node [draw=none] at (9,-0.25) {o};

\node [draw=none, mygreen] at (1.5,0.5) {$\three$};
\node [draw=none, myblue] at (4.5,0.5) {$\two$};
\node [draw=none, myred] at (7.5,0.5) {$\one$};
\end{tikzpicture}
\caption{Sketch of light propagation within three different spacetime geometries. The observer is located at region $\one$. The source is located at region $\three$. The points $h$ and $g$ are the identifiers, on the central null ray, of the boundaries between regions $\one - \two$ and $\two -\three$ respectively.}\label{fig:three_geom}
\end{figure}
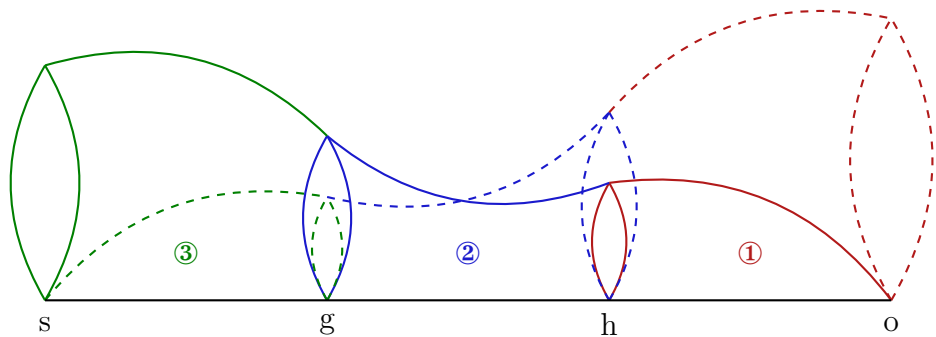

The discussions above are relevant for light propagation within a single spacetime between a vertex { {(observation)}} point and a source. Let us now consider light propagation within a universe that cannot be modeled by a single geometry. As an example, consider light propagation between three regions, \Rone, \Rtwo and \Rthree  which are modeled by different spacetime metrics that are not isometric to each other. We locate our observer in \Rone and the source in \Rthree with an arbitrary intervening region \Rtwo. (See figure~\ref{fig:three_geom}.) 
Then, in order to find the distance between the observer and the source, one has to propagate  {light throughout multiple geometries and solve for the Jacobi fields with non-vertex initial conditions within each region.}

{In fact, this was investigated by Fleury \textit{et al.} in \citep{Fleury:2013sna} in order to calculate distances in a Swiss-cheese universe. The authors consider a Wronski matrix ($\mathcal{W}$) method to solve the first order geodesic deviation equation throughout the propagation of light in the cheese and in the holes. Note that their Wronski matrix is exactly equal to our transfer matrix $\mathbf{T}$ when our Hamiltonian equations (\ref{eq:four_sets}) are imposed. The usefulness of this method was proven in many applications including  \citep{Fleury:2015rwa},  \citep{Fleury:2014gha} and  \citep{Fleury:2014rea}.}

 {We already mentioned that one needs to satisfy condition (\ref{eq:Recip_B}) in order for the distance reciprocity to hold within any setting. However, this is not immediately obvious for our example as}
\begin{eqnarray}
\mathbf{B}\left(v_s,v_o\right)&\neq & \mathbf{\Bthree}\left(v_s,v_g\right)\mathbf{\Btwo}\left(v_g,v_h\right)\mathbf{\Bone}\left(v_h,v_o\right),\nonumber \\
\mathbf{B}\left(v_o,v_s\right)&\neq & \mathbf{\Bone}\left(v_o,v_h\right)\mathbf{\Btwo}\left(v_h,v_g\right)\mathbf{\Bthree}\left(v_g,v_s\right).
\end{eqnarray}
 {Rather, one needs to (i) obtain the overall ray bundle transfer matrices $\mathbf{T}\left(v_s,v_o\right)$ and $\mathbf{T}\left(v_o,v_s\right)$ in both directions, (ii) read off the upper right corners of those transfer matrices in order to check whether the condition (\ref{eq:Recip_B}) holds or not.}

 {Note that, for light propagation from $o$ to $s$, one can define the overall ray bundle transfer matrix as 
\begin{eqnarray}\label{eq:Transfer_three_reg}
\mathbf{T}\left(v_s,v_o\right)
&=&\mathbf{\Tthree}\left(v_s,v_g\right)\mathbf{\Ttwo}\left(v_g,v_h\right)\mathbf{\Tone}\left(v_h,v_o\right)\nonumber \\
\left[
\begin{array}{c|c}
\mathbf{A} & \mathbf{B} \\
\hline
\mathbf{C} & \mathbf{D}
\end{array}
\right]
&=&
\left[
\begin{array}{c|c}
\mathbf{\Athree} & \mathbf{\Bthree} \\
\hline
\mathbf{\Cthree} & \mathbf{\Dthree}
\end{array}
\right]
\left[
\begin{array}{c|c}
\mathbf{\Atwo} & \mathbf{\Btwo} \\
\hline
\mathbf{\Ctwo} & \mathbf{\Dtwo}
\end{array}
\right]
\left[
\begin{array}{c|c}
\mathbf{\Aone} & \mathbf{\Bone} \\
\hline
\mathbf{\Cone} & \mathbf{\Done}
\end{array}
\right]
\end{eqnarray}
This follows from the fact that the space of 4-dimensional symplectic matrices, $Sp(4,\mathbb{R})$, forms a group under matrix multiplication. Thus, multiplication of two symplectic matrices is another symplectic matrix. Therefore, it is natural to expect the symmetry (\ref{eq:Recip_B}) to hold for the entire light propagation as the fact that $\mathbf{B}\left(v_{{f}},v_{{i}}\right)=-\mathbf{B}^{\intercal}\left(v_{{i}},v_{{f}}\right)
$ is true for any initial and final points follows from the symplectic symmetries of the transfer matrix. We will demonstrate this in the next subsection in more detail.}  

 {\subsection{Reciprocity and symplectic symmetries}}

 {Consider the composition map given in (\ref{eq:Transfer_three_reg}) for light propagation from point $o$ to $s$. Reading the upper right corner of the overall transfer matrix gives}
\begin{align}\label{eq:B_comp_so}
\mathbf{B}\left(v_s,v_o\right)&=\Athree \left(v_s,v_g\right)\Bigl[\Atwo \left(v_g,v_h\right)\Bone\left(v_h,v_o\right)
+\Btwo\left(v_g,v_h\right)\Done\left(v_h,v_o\right) \Bigr]  \nonumber \\
&+ \Bthree\left(v_s,v_g\right)\Bigl[\Ctwo\left(v_g,v_h\right)\Bone\left(v_h,v_o\right)
+\Dtwo\left(v_g,v_h\right)\Done\left(v_h,v_o\right)\Bigr].
\end{align}
Likewise for light propagation from point $s$ to $o$,
\begin{align}\label{eq:B_comp_os}
\mathbf{B}\left(v_o,v_s\right)&=\Aone \left(v_o,v_h\right)\Bigl[\Atwo \left(v_h,v_g\right)\Bthree\left(v_g,v_s\right)
+\Btwo\left(v_h,v_g\right)\Dthree \left(v_g,v_s\right)\Bigr] \nonumber \\
&+\Bone\left(v_o,v_h\right)\Bigl[\Ctwo\left(v_h,v_g\right)\Bthree\left(v_g,v_s\right)
+\Dtwo\left(v_h,v_g\right)\Dthree\left(v_g,v_s\right)\Bigr].
\end{align}
Then one raises the question: Under which conditions {the matrix (\ref{eq:B_comp_so}) is equal to minus transpose of the matrix given in (\ref{eq:B_comp_os})}, so that the distance reciprocity is satisfied?

Note that every symplectic matrix $\mathbf{T}$ has an inverse
\begin{eqnarray}
\mathbf{T}^{-1}=\mathbf{\Omega}^{-1}\,\mathbf{T}^{\intercal}\mathbf{\Omega},
\end{eqnarray}
 {which is also symplectic. This corresponds to}
\begin{eqnarray}\label{eq:Inv_symplectic}
\left[
\begin{array}{c|c}
\mathbf{A} & \mathbf{B} \\
\hline
\mathbf{C} & \mathbf{D}
\end{array}
\right]^{-1}\left(v_f,v_i\right)=\left[
\begin{array}{c|c}
\mathbf{D^{\intercal}} & \mathbf{-B^{\intercal}} \\
\hline
\mathbf{-C^{\intercal}} & \mathbf{A^{\intercal}}
\end{array}
\right]\left(v_f,v_i\right).
\end{eqnarray}
Moreover, as $\mathbf{T}\left(v_f,v_i\right)$ takes $\mathbf{z _{i}}$ set of solutions to $\mathbf{z _{f}}$ by a symplectic transformation, its inverse should take $\mathbf{z _{f}}$ to $\mathbf{z _{i}}$ for any initial and final point, i.e.,
\begin{eqnarray}\label{eq:Inv_transform}
\left[
\begin{array}{c|c}
\mathbf{A} & \mathbf{B} \\
\hline
\mathbf{C} & \mathbf{D}
\end{array}
\right]^{-1}\left(v_f,v_i\right)=\left[
\begin{array}{c|c}
\mathbf{A} & \mathbf{B} \\
\hline
\mathbf{C} & \mathbf{D}
\end{array}
\right]\left(v_i,v_f\right).
\end{eqnarray}
Then, through (\ref{eq:Inv_symplectic}) and (\ref{eq:Inv_transform}) we have
\begin{eqnarray}\label{eq:recip_matrices}
\mathbf{A}\left(v_i,v_f\right)&=\mathbf{D^{\intercal}}\left(v_f,v_i\right),\nonumber \\
\mathbf{B}\left(v_i,v_f\right)&=\mathbf{-B^{\intercal}}\left(v_f,v_i\right),\nonumber \\
\mathbf{C}\left(v_i,v_f\right)&=\mathbf{-C^{\intercal}}\left(v_f,v_i\right),\nonumber \\
\mathbf{D}\left(v_i,v_f\right)&=\mathbf{A^{\intercal}}\left(v_f,v_i\right).
\end{eqnarray}
 {Then, for the equations (\ref{eq:B_comp_so}) and (\ref{eq:B_comp_os}), one can show that $\mathbf{B}\left(v_s,v_o\right)=-\mathbf{B}^T\left(v_o,v_s\right)$ is true either (i) on account of the second relation of the equation set (\ref{eq:recip_matrices}) holding for any initial and final points or (ii) by making use of the equation set (\ref{eq:recip_matrices}) within each region.} Then the distance reciprocity (\ref{eq:Etherington}) is satisfied. This is true for light propagation through arbitrary number of regions, each region being modeled by an arbitrary spacetime. The only restriction we have is the continuity of the phase space vector throughout its evolution.

We would also like to emphasize the fact that (\ref{eq:Inv_symplectic}) also imposes certain symmetry conditions on submatrices. Namely, treating $\mathbf{T}$ as any block matrix on the left hand side of (\ref{eq:Inv_symplectic}) and taking its inverse gives
\begin{eqnarray}\label{eq:Block_inv}
\mathbf{D}^{\intercal}&=\left(\mathbf{A}-\mathbf{B}\mathbf{D}^{-1}\mathbf{C}\right)^{-1}\nonumber \\
-\mathbf{B}^{\intercal}&=-\left(\mathbf{A}-\mathbf{B}\mathbf{D}^{-1}\mathbf{C}\right)^{-1}\mathbf{B}\mathbf{D}^{-1}\nonumber \\
-\mathbf{C}^{\intercal}&=-\mathbf{D}^{-1}\mathbf{C}\left(\mathbf{A}-\mathbf{B}\mathbf{D}^{-1}\mathbf{C}\right)^{-1}\nonumber \\
\mathbf{A}^{\intercal}&=\mathbf{D}^{-1}+\mathbf{D}^{-1}\mathbf{C}\left(\mathbf{A}-\mathbf{B}\mathbf{D}^{-1}\mathbf{C}\right)^{-1}\mathbf{B}\mathbf{D}^{-1}.
\end{eqnarray}
Those constraints are equivalent to the so called \textit{symplectic conditions} in the literature which are given by
\begin{eqnarray}
&\mathbf{A}\mathbf{B}^{\intercal},\,\mathbf{A}^{\intercal}\mathbf{\dot{A}},\,\mathbf{B}^{\intercal}\mathbf{\dot{B}}\,\,\rm{and}\,\, \mathbf{\dot{A}}\mathbf{\dot{B}}^{\intercal}\,\, \rm{are\,\,symmetric},\nonumber \\
&\mathbf{A}\mathbf{\dot{B}}^{\intercal}-\mathbf{B}\mathbf{\dot{A}}^{\intercal}=\mathbf{I_2},\label{eq:symm_matrices}
\end{eqnarray}
when we impose the ray bundle transfer matrix evolution equations (\ref{eq:four_sets}). Indeed, it is easy to check that those follow from the very definition of a symplectic matrix given in (\ref{eq:Symplectic_T}). In our case, those symmetries essentially follow from the symmetries of the Riemann curvature tensor. {For a generic curvature tensor which includes the torsion term, neither the first order action (\ref{eq:The_action}) nor the reduced Lagrangian (\ref{eq:Red_Lagrangian}) takes the functional form we presented in this work. Accordingly, the equations of motion for generic deviation of curves are more involved \citep{Swaminarayan:1983}. Therefore, for alternative gravitation theories, that include torsion, the concept of distance reciprocity is questionable.}

Thus, we conclude that the distance reciprocity in {general} relativity follows from the symplectic symmetries of the underlying first order light propagation system. 

\section{Canonical transformations and generating functions}\label{Canonical transformations and generating functions}

Our ray bundle transfer matrices create linear symplectomorphisms on the phase space which are known as linear canonical transformations in physics. Canonical transformations preserve the form of the Hamiltonian equations by leaving the Poisson bracket invariant up to a constant. Then it is natural to look for the generating function of this transformation.

For our canonical transformation $f:\mathbb{R}^4\rightarrow \mathbb{R}^4$ with
\begin{eqnarray}
\boldsymbol{\xi '} \rightarrow \boldsymbol{\xi}=\boldsymbol{\xi}(\boldsymbol{\xi '},\boldsymbol{\dot{\xi '}};v),\label{eq:final_xi} \\
\boldsymbol{\dot{\xi '}} \rightarrow \boldsymbol{\dot{\xi}}=\boldsymbol{\dot{\xi}}(\boldsymbol{\xi '},\boldsymbol{\dot{\xi}'};v),\label{eq:final_xidot}
\end{eqnarray}
there exists an associated 1-form
\begin{eqnarray}\label{eq:1form_dS}
d\tilde{S}(\boldsymbol{\dot{\xi}'},\boldsymbol{\xi '};v)=\boldsymbol{\dot{\xi}'}d\boldsymbol{\xi '}-\boldsymbol{\dot{\xi}}d\boldsymbol{\xi},
\end{eqnarray}
which is exact.

For the time being, we are interested in those transformations in which
\begin{eqnarray}\label{eq:free_can_trans}
{\rm{det}}\frac{\partial\left(\boldsymbol{\xi},\boldsymbol{\xi '}\right)}{\partial \left(\boldsymbol{\dot{\xi '}},\boldsymbol{\xi '}\right)}={\rm{det}}\frac{\partial \boldsymbol{\xi}}{\partial \boldsymbol{\dot{\xi}'}}={\rm{det}}\mathbf{B}\neq 0,
\end{eqnarray}
so that the angular diameter and luminosity distances given in (\ref{eq:D_A_D_L_B})
can be computed. A transformation characterized by the condition (\ref{eq:free_can_trans}) is known as a \textit{free canonical transformation} in the literature \citep{Arnold:1978}. In this case, the function $\tilde{S}$ can be locally expressed as 
\begin{eqnarray}
\tilde{S}(\boldsymbol{\dot{\xi}},\boldsymbol{\xi};v)=S(\boldsymbol{\xi},\boldsymbol{\xi '};v),
\end{eqnarray}
with $S(\boldsymbol{\xi},\boldsymbol{\xi '};v)$ being the generating function of our free canonical transformation. It is given by
\begin{eqnarray}
S(\boldsymbol{\xi},\boldsymbol{\xi '};v)=\int _{\boldsymbol{\xi '}, 0}^{\boldsymbol{\xi }, v}\boldsymbol{\dot{\xi}}d\boldsymbol{\xi}-Hdv,
\end{eqnarray}
and is equal to our quadratic geodesic deviation action, (\ref{eq:The_action}), derived via Synge's world function.

For a linear, free canonical transformation, represented by a symplectic ABCD block matrix, one can write $S(\boldsymbol{\xi},\boldsymbol{\xi'};v)$ by matrix inner products \citep{deGosson:2006}
\begin{eqnarray}\label{eq:gen_fun}
S=\frac{1}{2}(\mathbf{D}\mathbf{B}^{-1}\boldsymbol{\xi},\boldsymbol{\xi})-(\mathbf{B}^{-1}\boldsymbol{\xi},\boldsymbol{\xi'})
+\frac{1}{2}(\mathbf{B}^{-1}\mathbf{A}\boldsymbol{\xi'},\boldsymbol{\xi'}).
\end{eqnarray}
In the Appendix, we show that $S(\boldsymbol{\xi},\boldsymbol{\xi '};v)$ satisfies
\begin{eqnarray}\label{eq:trans_evol}
\boldsymbol{\dot{\xi}}=\frac{\partial S}{\partial \boldsymbol{\xi}}, \,\,\, \boldsymbol{\dot{\xi}'}=-\frac{\partial S}{\partial \boldsymbol{\xi}'},\,\,\,
\frac{\partial S}{\partial v}+H=0,
\end{eqnarray}
as one would expect from a generating function of a free canonical transformation \citep{deGosson:2006}. We should note that writing our geodesic deviation action $S(\boldsymbol{\xi},\boldsymbol{\xi '};v)$ in the form of (\ref{eq:gen_fun}) is paramount for switching back to wave optics picture from paraxial ray bundles as we discuss in Section~\ref{Wavization of a ray bundle}.
\section{Density function and Liouville's equation}\label{Density function and Liouville's equation}
Our 4-dimensional symplectic phase space is endowed with a volume element
\begin{eqnarray}
d\mathcal{V}=d\xi^1\wedge d\xi^2 \wedge d\dot{\xi}_1 \wedge d\dot{\xi}_2.
\end{eqnarray}
Accordingly, we define the total number of light rays within the bundle as
\begin{eqnarray}\label{eq:phase_sp_density}
\mathcal{N}=\int_{\mathcal{V}} \mathcal{n}(\boldsymbol{\xi},\boldsymbol{\dot{\xi}};v)d\mathcal{V},
\end{eqnarray}
in which $\mathcal{n}(\boldsymbol{\xi},\boldsymbol{\dot{\xi}};v)$ is the phase space density function, i.e., number of photons per unit phase space volume.

The phase space volume element is an invariant of the symplectic phase space. This follows from the invariance of the underlying symplectic structure\footnote{For more details see our accompanying paper  \citep{Uzun:2018} which, in addition, focuses on invariance of phase space volume under some \textit{virtual} Hamiltonian flow to prove Etherington's distance reciprocity in an abstract form.} \citep{Arnold:1978} . Moreover, \textit{if} we have a lossless/gainless system then the number of light rays piercing the observational screen is conserved. In that case, the phase space density is invariant throughout the evolution, with respect to the affine parameter of the null geodesic. Then, the Liouville equation is as follows \citep{Torre:2005}
\begin{eqnarray}
\frac{d\mathcal{n}(\boldsymbol{\xi},\boldsymbol{\dot{\xi}};v)}{dv}&=&0\nonumber \\
&=&\frac{\partial \mathcal{n}}{\partial v}+\frac{\partial \mathcal{n}}{\partial \boldsymbol{\xi}}\frac{d \boldsymbol{\xi}}{d v}+\frac{\partial \mathcal{n}}{\partial \boldsymbol{\dot{\xi}}}\frac{d \boldsymbol{\dot{\xi}}}{d v}\nonumber \\
&=&\frac{\partial \mathcal{n}}{\partial v}+\frac{\partial \mathcal{n}}{\partial \boldsymbol{\xi}}\frac{\partial H}{\partial \boldsymbol{\dot{\xi}}}+\frac{\partial \mathcal{n}}{\partial \boldsymbol{\dot{\xi}}}\left(\frac{-\partial H}{\partial \boldsymbol{\xi}}\right)\nonumber \\
&=&\frac{\partial \mathcal{n}}{\partial v}+\left\{H,\mathcal{n}\right\},
\end{eqnarray}
in which the third line follows from the Hamilton's equation, (\ref{eq:Hamilton_Jacobi_z}); and the fourth line from the definition of the Poisson bracket, (\ref{eq:Poisson_funcs}). Then, we can simply write 
\begin{eqnarray}
\frac{\partial \mathcal{n}}{\partial v}=\hat{\mathcal{L}}_H[\mathcal{n}],
\end{eqnarray}
where $\hat{\mathcal{L}}_H[\bullet]$ is the Lie operator defined in (\ref{eq:Lie_op}).
\section{Potential applications of reduced phase space optics}\label{Potential applications of reduced phase space optics}
In the Newtonian case, applications of symplectic phase space optics for ABCD systems are vast. In this section, we will briefly \textit{sketch} the potential applications of our construction relevant for cosmological and astrophysical observations. 
\subsection{Phase space averaging}
As light propagates within the universe it carries information about the \textit{averaged} footprints of the phenomena that affect its propagation. Some of those footprints are assumed to cancel out throughout the propagation in the standard, perturbative scheme of standard cosmology. 

On the other hand, inhomogeneous cosmological models have become more popular than ever over the past few decades. Recognition of the fact that the inhomogeneities in the universe might not average out to define a spatially flat universe at late times  {\citep{Ellis:1984}}, lead researchers in this field to address the following question. Can the late time inhomogenities in the universe be responsible for (at least some portion of) the apparent accelerated expansion of the universe, rather than the so called dark energy?

Accordingly, averaging techniques on spatial hypersurfaces \citep{Buchert:2000,Buchert:2001} have been investigated in many papers \citep{Rasanen:2004, Wiltshire:2007, Wiltshire:2009, Rasanen:2009, Mattsson:2010, Kolb:2011, Ellis:2011, Buchert:2012, Bagheri:2014} to determine their consequences on cosmological distances and the Hubble parameter. In some of these works, it is assumed that light propagates on a spacetime with smoothed out 3-dimensional spatial hypersurfaces, effectively. 
Thus, the main idea is to study the effect of light propagation via the averaging of the 3-dimensional \textit{configuration space}. The hypersurface average of a function $f(x^\mu)$ on a spatial domain, $D$, is given by
\begin{eqnarray}\label{eq:spatial_average}
\langle f \rangle_D(x^\mu)=\frac{\int_Df(x^\mu)\sqrt{h_{ij}}d^3x}{\int_D\sqrt{h_{ij}}d^3x},
\end{eqnarray}
in which $\sqrt{h_{ij}}$ corresponds to the {square root of the} determinant of the 3 - metric induced on the spatial hypersurfaces and $d^3x$ is the coordinate volume element. Accordingly, the denominator of (\ref{eq:spatial_average}) can be interpreted as the spatial (proper) volume of the domain measured by the observers depending on their foliation 4-velocity. Note that the averaged dynamics is then foliation dependent \citep{Buchert:2018}.

On the other hand, the necessity of null cone averages in cosmology have been discussed by many authors \citep{Dyer:1988, Coley:2009, Marra:2007, Gasperini:2011} as the observables are averaged via the propagation of light, not over the spatial domains. Accordingly, we propose an alternative, covariant averaging method on our reduced phase space. Consider the following classical phase space average of a function $f(\boldsymbol{\xi},\dot{\boldsymbol{\xi}};v)$
\begin{eqnarray}\label{eq:cl_phase_sp_ave}
\bar{f}(\boldsymbol{\xi},\dot{\boldsymbol{\xi}};v)=\frac{1}{\mathcal{N}}\int f(\boldsymbol{\xi},\dot{\boldsymbol{\xi}};v)\mathcal{n}(\boldsymbol{\xi},\dot{\boldsymbol{\xi}};v)d\boldsymbol{\xi}d\dot{\boldsymbol{\xi}},
\end{eqnarray}
where $\mathcal{n}(\boldsymbol{\xi},\dot{\boldsymbol{\xi}};v)$ is the phase space distribution function, namely the number density function defined through (\ref{eq:phase_sp_density}). For a lossless/gainless system $\mathcal{n}(\boldsymbol{\xi},\dot{\boldsymbol{\xi}};v)$ is conserved due to Liouville's theorem as outlined in Section~\ref{Density function and Liouville's equation}. Then the commutation relation between the evolution operator and the phase space averaging follows as 
\begin{eqnarray}
\frac{d\bar{f}(\boldsymbol{\xi},\dot{\boldsymbol{\xi}};v)}{dv}=\overline{\frac{df(\boldsymbol{\xi},\dot{\boldsymbol{\xi}};v)}{dv}}.
\end{eqnarray}
This indicates that averaging of a scalar on the phase space commutes with its evolution with respect to the affine parameter $v$.

Furthermore, we can consider a $v$-average of the ensemble average of a function to get
\begin{eqnarray}
f_{\vee}=\frac{\int\bar{f}(\boldsymbol{\xi},\dot{\boldsymbol{\xi}};v)\,dv}{\int dv},
\end{eqnarray}
which gives us a \textit{null bundle average} of the function in question.
 {Note that for most of the scalar averaging techniques within a $3+1$ decomposition of the spacetime, the evolution and constraint equations have three main ingredients : (i) the Hamiltonian constraint, (ii) scalar
projection of the energy-momentum conservation equation and (iii) contracted Raychaudhuri equation for a timelike worldline. Therefore, an observed null bundle average could indeed be more promising as it allows one to average, the \textit{full} set of Einstein equations in principle, say, under the spin field formalism of Newman and Penrose \citep{Newman:1961}.}

\subsection{Spacetime \texorpdfstring{$\equiv$}{} thin lens, pure magnifier and fractional Fourier transformer}
Any symplectic matrix belonging to $Sp(2,\mathbb{R})$ can be decomposed uniquely into three matrices that belong to a maximally compact subgroup, an abelian subgroup and a nilpotent subgroup. Such a decomposition is named after Iwasawa \citep{Iwasawa:1949}. This fact is used in Newtonian optics for a system whose ray transfer is given by a symplectic matrix such that the optical system can be decomposed into a fractional Fourier transformer, a pure magnifier and a thin lens \citep{Simon:1998}. 

For symplectic matrices that belong to $Sp(4,\mathbb{R})$, as in our case, one defines a modified Iwasawa factorization\footnote{In higher dimensions this is a factorization, i.e., a parameterization of the group, rather than a decomposition. This is due to the fact that symmetric matrices which appear in the pure magnifier component do not form a group under multiplication \citep{Wolf:2004}.} as the following \citep{Sundar:1995, Simon:2000, Wolf:2004}
\begin{eqnarray}\label{eq:Iwasawa}
\left[
\begin{array}{c|c}
\mathbf{A} & \mathbf{B} \\
\hline
\mathbf{C} & \mathbf{D}
\end{array}
\right]&=&\left[
\begin{array}{c|c}
\mathbf{I_2} & \mathbf{0_2} \\
\hline
\mathbf{-G} & \mathbf{I_2}
\end{array}
\right]
\left[
\begin{array}{c|c}
\mathbf{S} & \mathbf{0_2} \\
\hline
\mathbf{0_2} & \mathbf{S^{-1}}
\end{array}
\right]
\left[
\begin{array}{c|c}
\rm{Re}\mathbf{U} & \rm{Im}\mathbf{U} \\
\hline
-\rm{Im}\mathbf{U} & \rm{Re}\mathbf{U}
\end{array}
\right]\nonumber\\
\nonumber\\
&=&\qquad\mathbf{L(G)}\qquad \,\,\mathbf{M(S)}\qquad \qquad\mathbf{F(U)}\nonumber \\
&=&\,\,\,\,\,\,\,\,\,\,
\begin{array}{c}
Thin\\
lens
\end{array}\,\,\,\,
\begin{array}{c}
Pure\\
magnifier
\end{array}\,\,\,
\begin{array}{c}
Fractional\\
Fourier\\
transformer
\end{array}\nonumber\\
\end{eqnarray}
Here the $2\times2$ matrices that appear in (\ref{eq:Iwasawa}) are given by
\begin{eqnarray}
\mathbf{G}&=&-\left(\mathbf{\dot{A}}\mathbf{A}^{\intercal}+\mathbf{\dot{B}}\mathbf{B}^{\intercal}\right)\left(\mathbf{A}\mathbf{A}^{\intercal}+\mathbf{B}\mathbf{B}^{\intercal}\right)^{-1}=\mathbf{G}^{\intercal}\nonumber\\
\mathbf{S}&=&\left(\mathbf{A}\mathbf{A}^{\intercal}+\mathbf{B}\mathbf{B}^{\intercal}\right)^{1/2}=\mathbf{S}^{\intercal}\nonumber\\
\mathbf{U}&=&\left(\mathbf{A}\mathbf{A}^{\intercal}+\mathbf{B}\mathbf{B}^{\intercal}\right)^{-1/2}\left(\mathbf{A}+i\mathbf{B}\right)\in U(2),
\end{eqnarray}
once we impose the  {thin} ray bundle evolution equations, (\ref{eq:four_sets}). This means that light propagation in \textit{any} spacetime between \textit{any} initial and final points can be uniquely factored into its thin lens, pure magnifier and fractional Fourier transformer components. The thin lens component is responsible for a shearing effect in the $\boldsymbol{\dot{\xi}}$ direction on the phase space. The matrix $\mathbf{S}$ provides a magnification in $\boldsymbol{{\xi}}$ direction and a demagnification in $\boldsymbol{\dot{\xi}}$. The fractional Fourier component \citep{Mendlovic:1993}, on the other hand, is a generalization of phase space rotations \citep{Simon:2000b}. 

In particular, consider our canonical pairs $\{\boldsymbol{{\xi}},\boldsymbol{\dot{\xi}}\}$ to be ordinary Fourier pairs. Then an ordinary integral Fourier transform can be written which takes a function, $f(\boldsymbol{\xi};v)$, in  a $\boldsymbol{\xi}$ domain to a function, $\tilde{f}(\boldsymbol{\dot{\xi}};v)$ in a $\boldsymbol{\dot{\xi}}$ domain by 
\begin{eqnarray}
\tilde{f}(\boldsymbol{\dot{\xi}};v)=\int f(\boldsymbol{\xi};v)\,\exp{[{-2\pi i \boldsymbol{\dot{\xi}}\cdot \boldsymbol{\xi}}]} d\boldsymbol{\xi}.
\end{eqnarray}  
Indeed, such a transformation takes $\boldsymbol{\xi}\rightarrow \,\boldsymbol{\dot{\xi}}$ and $\boldsymbol{\dot{\xi}}\rightarrow \,-\boldsymbol{\xi}$. Its discreet version is given by a specific form of the generalized matrix $\mathbf{F(U)}$ in (\ref{eq:Iwasawa}), i.e., when $\mathbf{U}=i\mathbf{I_2}$\footnote{In a 2-dimensional phase space, we have $U=i$ and the ordinary Fourier transform corresponds to a $\pi/2$ rotation of the phase space coordinates.}.

The fractional Fourier transformation, being a generalization of the ordinary Fourier transform, serves as an important tool in the Newtonian wave optics \citep{Ozaktas:1999}. The analysis of the transformation of the quasiprobability distribution of the wavized phase space is closely related to fractional Fourier transformations. Accordingly, it can serve as a means to identify whether Gaussian wave packets remain Gaussian \citep{Simon:1998} throughout the propagation in a given spacetime. Moreover, fractional Fourier transforms are important for the phase space tomography techniques of the Newtonian theory \citep{Alieva:2016} in which the intervening optical system properties are \textit{derived} in an inverse problem. It is an interesting, open question whether or not such a spacetime tomography method can be developed for segmented portions of our line of sight, given the initial and final forms of our phase space vector $\mathbf{z}$ at each point.

\subsection{Wavization of a ray bundle}\label{Wavization of a ray bundle}
Even though astrophysical objects are too large for the wave effects to be observed and that the ray picture is a good approximation for many applications, wave optics is still relevant for many areas in relativity. For instance, polarization optics is important for extraction of cosmological parameters via the cosmic microwave background (CMB) radiation. Likewise, polarization of the radio emission of pulsars and active galaxies are important for extraction of properties of the interstellar medium, emission processes, etc. Detection of black holes via their shadows is well within the wave optics regime as the apparent sizes of the shadows are very small and diffraction effects are crucial for their identification. 

Note that just as classical mechanics agree with quantum mechanics in the $\hbar \rightarrow 0$ limit for linear systems; geometric optics agree with wave optics in the small wave length limit, up to first order. In order to recover wave optics from the ray picture in the paraxial regime, however, one needs to use certain quantization techniques \citep{Raszillier:1986, Castanos:1986}. Such an argument follows from the analogy between quantum mechanics and classical paraxial optics. The phase space of classical mechanics is the one of the geometric optics and the phase space of quantum mechanics is the same as the one of the wave optics for first order light propagation \citep{Dragoman:2002}. 

Let us be more specific. It is known that the symplectic group $Sp(2n,\mathbb{R)}$ has a unique double cover known as the \textit{metaplectic group}, $Mp(2n,\mathbb{R)}$ \citep{Simon:2000, Wolf:2004}. Accordingly, linear canonical transformations have unitary representations \citep{Healy:2015}. Following this, the idea of wavization of an \textit{observed} ray pencil of a curved background should in principle follow from finding operator representations of our projected Jacobi fields and their derivatives, namely, 
\begin{eqnarray}
\boldsymbol{\xi}\rightarrow \boldsymbol{\hat{{\xi}}} \qquad \boldsymbol{\dot{\xi}}\rightarrow \boldsymbol{\hat{\dot{{\xi}}}}.
\end{eqnarray}
The evolution of the operators can be then given by similarity transformations,
\begin{eqnarray}
\boldsymbol{\hat{{\xi}}}(v)&=&\mathbf{\hat{{T}}}^{-1}(v,v_i)\boldsymbol{\hat{{\xi}}}(v_i)\mathbf{\hat{{T}}}(v,v_i),\nonumber \\
\boldsymbol{\hat{\dot{{\xi}}}}(v)&=&\mathbf{\hat{{T}}}^{-1}(v,v_i)\boldsymbol{\hat{\dot{{\xi}}}}(v_i)\mathbf{\hat{{T}}}(v,v_i).
\end{eqnarray}
Here, $\mathbf{\hat{{T}}}(v,v_i)$ is the unitary operator associated with the ray bundle transfer matrix $\mathbf{T}(v,v_i)$, (\ref{eq:Transfer_OE}). Now, let us consider the plane that is transverse to the null vector $\vec{k}$ and spanned by the Sachs basis components of $\vec{\xi}$. Instead of covering the full solutions of Maxwell's equations on a given spacetime, one can consider only the transverse components of electromagnetic wave function that are most relevant for the observations. Those solutions would be then analogous to the parabolic wave equations of Newtonian optics. These are the approximate solutions of spherical wave functions in the paraxial regime  \citep{Torre:2005}. Then the initial transverse electromagnetic wave function, or the complex amplitude, $E(\boldsymbol{\xi'};v')$, is advanced to a final complex amplitude by $E(\boldsymbol{\xi};v)=\mathbf{\hat{{T}}}(v,v')E(\boldsymbol{\xi'};v')$. Following Moshinsky and Quesne's seminal work \citep{Moshinsky:1971,Liberman:2015}, such a transformation can be written as an integral transform \citep{Torre:2005, Alonso:2011, Wolf2016}
\begin{eqnarray}\label{eq:diffraction_integral}
E(\boldsymbol{\xi};v)=\int \frac{1}{\sqrt{(2\pi i)^2{\rm{det}}\mathbf{B}}}K\left(\boldsymbol{\xi},\boldsymbol{\xi'};v\right)E(\boldsymbol{\xi'};v')d\boldsymbol{\xi'},
\end{eqnarray}
for each component of the transverse wave function aligned with the Sachs basis. Here, the kernel $K\left(\boldsymbol{\xi},\boldsymbol{\xi'};v\right)$ is given by
\begin{eqnarray}
K\left(\boldsymbol{\xi},\boldsymbol{\xi'};v\right)=\exp{[{iS(\boldsymbol{\xi},\boldsymbol{\xi '};v)}]},
\end{eqnarray}
and $S(\boldsymbol{\xi},\boldsymbol{\xi '};v)$ is given by (\ref{eq:gen_fun}), which is the generating function of the underlying free canonical transformation. It is written in terms of the elements of the transfer matrix $\mathbf{T}$ as we discussed in Section~\ref{Canonical transformations and generating functions}. Then equation (\ref{eq:diffraction_integral}) is the most generic form of Huygens diffraction integral which is also known as Collins integral \citep{Collins:1970}, for first order light propagation in a given spacetime. Note that in Section~\ref{Canonical transformations and generating functions} we mentioned that the generating function $S(\boldsymbol{\xi},\boldsymbol{\xi '};v)$, which shows up as a phase factor here, is indeed our geodesic deviation action up to quadratic order. It defines  {$v =$} constant planes.  Here, in the wave picture, it serves a tool to identify stationary phase surfaces which analogously approximates Huygen's principle in paraxial wave optics.

We plan to elaborate on the wavization of an observed ray bundle in a forthcoming paper via a rigorous quantization technique. For the current section, all we wanted demonstrate are the links between our geodesic deviation action, the generating function of the corresponding linear symplectomorphism and the kernel of the wave function transformation when the ray bundle is quantized. This means that for light bundle propagation for first order optics, our quadratic deviation action is preeminent both for ray and wave optics pictures.
\subsection{Polarization optics and its evolution}
In Newtonian optics, there has been much work to investigate how the polarization state of a light beam changes as it passes through a generic first order ABCD system. Those investigations can indeed shed light on certain problems relevant for astrophysics and cosmology.

Given a $3+1$ decomposition of the spacetime in general relativity, the optical phase space is 6-dimensional with a volume element $d^3\mathbf{x}\,d^3\mathbf{p}$ where $\mathbf{x}$ are the induced spatial coordinates of the underlying spacetime metric and $\mathbf{p}$ are the 3-momenta of the photon. In cosmology, for example, one considers a polarization tensor to investigate the polarization states of the CMB radiation. However, it is the \textit{screen-projected linear} polarization tensor that is composed of the Stokes parameters, that incorporates the observable effects and which removes the residual gauge freedom in the problem \citep{Tsagas:2008, Ellis:2012}. Therefore, we advocate that a polarization matrix defined within our 4-dimensional reduced phase space would be as valuable as the screen-projected polarization tensor given in the literature. Let us now introduce the idea of a Wigner quasiprobability distribution in quantum mechanics, $W_{qn}$, which can then be used to construct a covariant polarization matrix. 

Wigner introduced a quantum mechanical analogue of the classical phase space density function in order to find the expectation values of operators on the phase space \citep{Wigner:1932}. For our canonical pair $\{\boldsymbol{\xi}, \dot{\boldsymbol{\xi}}\}$ such an ensemble average would look like
\begin{eqnarray}
\langle \hat{f} \rangle_{\substack{\\ \rm{qn.}}}=\int W_{\substack{\\ \rm{qn.}}}(\boldsymbol{\xi},\dot{\boldsymbol{\xi}};v)f(\boldsymbol{\xi},\dot{\boldsymbol{\xi}};v)d\boldsymbol{\xi}d\dot{\boldsymbol{\xi}}
\end{eqnarray}
which is analogous to the classical phase space average given in (\ref{eq:cl_phase_sp_ave}). Here, the correspondence between the operator $\hat{f}$ and the function $f$ is proposed by Weyl \citep{Weyl:1927} and shown by Moyal \citep{Moyal:1949}\footnote{See, for example, \citep{Hillery:1984} for a detailed review on the Wigner distribution in quantum mechanics.}.

The fact that a quantum mechanical quasiprobility distribution function is adopted by the classical optics community follows from the analogy between quantum mechanics and classical optics in the paraxial regime that we mentioned in the previous section. In this picture, mixed states of quantum mechanics are analogous to partially coherent light beams. Accordingly the density matrix that appears in the original definition of Wigner is replaced by a coherency matrix. Following this, an \textit{optical} Wigner distribution function was introduced into classical optics to study partially coherent light \citep{Walther:1968, Bastiaans:1978, Bastiaans:1979}. An optical Wigner matrix \citep{Duan:2005, Dragoman:2005, Castaneda:2006, Alonso:2011} can be written in our case as the following
\begin{eqnarray}
W_{ab}(\boldsymbol{\xi},\dot{\boldsymbol{\xi}};v)=\left(\frac{\kappa}{2\pi}\right)^2\int \Gamma _{ab}\left(\boldsymbol{\xi}-\frac{\boldsymbol{\xi '}}{2},\boldsymbol{\xi}+\frac{\boldsymbol{\xi '}}{2};v\right)\exp{[{i\kappa\boldsymbol{\xi '}\cdot \dot{\boldsymbol{\xi}}}]}d\boldsymbol{\xi '}
\end{eqnarray}
where 
\begin{eqnarray}\label{eq:coherence_mat}
\Gamma _{ab}(\boldsymbol{\xi _A},\boldsymbol{\xi _B};v)=\langle E_a(\boldsymbol{\xi _A};v), E_b(\boldsymbol{\xi _B};v)\rangle
\end{eqnarray}
is a $v$-dependent cross-spectral density matrix, $\{a,b\}=\{1,2\}$ refers to components of the field in the Sachs basis and $\kappa$ is a constant.

\textit{Generalized} Stokes parameters are then constructed from this optical Wigner 
matrix as the following \citep{Korotkova:2005, Luis:2006}
\begin{eqnarray}
S_0(\boldsymbol{\xi},\dot{\boldsymbol{\xi}};v)&=&W_{11}(\boldsymbol{\xi},\dot{\boldsymbol{\xi}};v)+W_{22}(\boldsymbol{\xi},\dot{\boldsymbol{\xi}};v),\nonumber \\
S_1(\boldsymbol{\xi},\dot{\boldsymbol{\xi}};v)&=&W_{11}(\boldsymbol{\xi},\dot{\boldsymbol{\xi}};v)-W_{22}(\boldsymbol{\xi},\dot{\boldsymbol{\xi}};v),\nonumber \\
S_2(\boldsymbol{\xi},\dot{\boldsymbol{\xi}};v)&=&W_{12}(\boldsymbol{\xi},\dot{\boldsymbol{\xi}};v)+W_{21}(\boldsymbol{\xi},\dot{\boldsymbol{\xi}};v),\nonumber \\
S_3(\boldsymbol{\xi},\dot{\boldsymbol{\xi}};v)&=&i\left[W_{12}(\boldsymbol{\xi},\dot{\boldsymbol{\xi}};v)-W_{21}(\boldsymbol{\xi},\dot{\boldsymbol{\xi}};v)\right].\nonumber \\
\end{eqnarray}
Note that these generalized Stokes parameters accommodate information about both of the Fourier pairs, i.e., position on the screen and frequency weighted direction. This makes its Poincar\'e sphere representation fairly simple \citep{Luis:2006}. Wigner matrix components are invariant throughout a symplectic ABCD propagation, i.e.,
\begin{eqnarray}
W_{ab}(\mathbf{A}\boldsymbol{\xi}+\mathbf{B}\dot{\boldsymbol{\xi}}, \mathbf{C}\boldsymbol{\xi}+\mathbf{D}\dot{\boldsymbol{\xi}} ;v)=W_{ab}(\boldsymbol{\xi},\dot{\boldsymbol{\xi}};v_0).
\end{eqnarray}
Therefore, the invariance applies for the Stokes vector, $\boldsymbol{s}=\left(S_0, S_1, S_2, S_3\right)^{\intercal}$ as well. This allows one to investigate the evolution of the two-point generalized Stokes vector $\boldsymbol{\tilde{s}}(\boldsymbol{\xi _A},\boldsymbol{\xi _B};v)$ \citep{Korotkova:2005, Du:2008, Sahin:2010} through out an ABCD system via Mueller matrices \citep{Luis:2006}. We believe such methodologies developed for optical devices in the Newtonian theory can be adopted to investigate the change of the polarization states of light beams in astrophysical and cosmological scenarios\footnote{Note that the polarization state corresponding to a \textit{single ray} is unaffected by the spacetime curvature as the components of the electromagnetic vector potential are parallely propagated with respect to the corresponding tangent vector of the null curve, in the geometric optics limit. However, we observe \textit{ray bundles} rather than individual rays. Therefore, it is natural to expect a change in the polarization state of an electromagnetic field \textit{with respect to} its fiducial null neighbor. This information should be carried by the corresponding geodesic deviation variables.}. For instance, curvature induced gravitational lensing or late time integrated Sachs-Wolfe effects on the polarization of the CMB can be examined in such a manner without introducing pertubative schemes within alternative cosmological models.
\subsection{Invariants, autonomization and stability analysis}
Following the early paper of Lewis on time-dependent harmonic oscillators in the classical theory \citep{Lewis:1967}, there has been considerable amount of work on finding the invariants and autonomization of nonautonomous systems \citep{Leach:1977,Gunther:1977,Lewis:1982,Kanasugi:1995,Struckmeier:2001,Struckmeier:2002,Struckmeier:2005}.

This is particularly of interest for the current investigation as our quadratic, oscillator-like Hamiltonian is also  a function of the evolution parameter, $v$. Note that this creates a technical difficulty in estimating the ordered exponentials in (\ref{eq:Transfer_OE}) in order to obtain the ray bundle propagation matrix $\mathbf{T}$. Moreover, the stability analysis of the observed light bundles under perturbations might be challenging as a well-defined theory of stability analysis exists either for linear autonomous or periodic systems only \citep{Meyer:2017}. Therefore, autonomization of our first order system is relevant for (i) reducing the ordered exponentials into simple Lie transformations and (ii) finding the answer to the question: under which conditions and in what kind of spacetimes, a ray bundle which is perturbed along its pathway diverges and ceases to be observable?  

One of the techniques of autonomization follows from extending the phase space of the physical system by two degrees of freedom. Consider the phase space coordinates, $\{\boldsymbol{\xi}, \dot{\boldsymbol{\xi}}\}$, evolution parameter, $v$, and the Hamiltonian $H(\boldsymbol{\xi}, \dot{\boldsymbol{\xi}}; v)$ of our system. Following the methodology outlined in  \citep{Struckmeier:2002} one can apply a $v$-dependent canonical transformation 
\begin{eqnarray}
(\boldsymbol{\xi}, \dot{\boldsymbol{\xi}}) \longrightarrow \left(\mathbf{q}, \mathbf{p}\right),
\end{eqnarray}
and reparameterize the evolution by
\begin{eqnarray}
v\rightarrow s,
\end{eqnarray}
such that a canonically equivalent system can be defined with a transformed, autonomous Hamiltonian function, $\tilde{H}$, 
\begin{eqnarray}
H(\boldsymbol{\xi}, \dot{\boldsymbol{\xi}}; v)\longrightarrow  \tilde{H}(\mathbf{q}, \mathbf{p}),
\end{eqnarray}
namely,
\begin{eqnarray}
\frac{1}{2}\tensor{\delta}{^{a}^{b}} \dot{\xi}_{a}\dot{\xi}_{b}-\frac{1}{2}\tensor{\mathcal{R}}{_{\,}_{a}_{b}}(v)\xi^{a}\xi^{b}  \longrightarrow  \frac{1}{2}\left[\mathbf{p}^2+V(\mathbf{q})\right],
\end{eqnarray}
in which $V(\mathbf{q})$ acts like a $s$-independent potential in the transformed system. The \textit{physical} equivalence of the original system and the transformed one then relies on the uniqueness of the underlying canonical transformation. 

Whether or not such a unique transformation can be found and a reparameterization of the evolution of the system, $v\rightarrow s$, can be obtained via an affine transformation, at least for certain types of spacetimes, is an interesting question not only for mathematical but also for physical aspects. Note that under the physical equivalence, $\tilde{H}(\mathbf{q}, \mathbf{p})$ corresponds to the invariant of the underlying system \citep{Struckmeier:2001, Struckmeier:2002}. Moreover, if one can write $V(\mathbf{q})$ as a quadratic function of $\mathbf{q}$ and follow the same procedure that we outlined in Section~\ref{Reduced Hamiltonian and ABCD Matrices}, then the propagation matrix is obtained via Lie transformations as we mentioned. This reduces the first order light propagation problem into a very simple form. Furthermore, the eigenvalues of the corresponding Hamiltonian matrix can be used to plot phase portraits and to obtain stability and bifurcation analysis. We leave these questions for further investigations.

 {As a final note, we would like to emphasize that even though the extended phase space technique seems to be promising, it is the symmetries of the underlying spacetime that simplifies the problem. The existence and uniqueness of the solutions can be investigated once the explicit form of the optical tidal matrix is obtained.}

\section{Summary and conclusion}\label{Summary and conclusion}
The use of symplectic methods in Newtonian optics became popular only after the 1980s. By that time, there were almost no open problems left in the general relativistic community in terms of light propagation in the geometric optics limit. On the other hand, after late the 1990s, the amount of cosmological data and the precision of experiments increased exponentially, revealing a highly inhomogeneous universe at late times. Whether or not the universe can be modeled by a unique geometry \citep{Fleury:2013uqa} or backreaction effects are significant for cosmological light propagation have been the subject of debate \citep{Buchert:2015}. Accordingly, light propagation for more complex, realistic scenarios and its effect on the observables have become of interest \citep{Fleury:2013sna,Bacon:2014,Korzynski:2017,Jolicoeur:2017}.

In this work, our aim was to construct the general relativistic analogue of the paraxial regime of Newtonian optics in a Machian setting. We believe that under such a construction, the improvements of the symplectic methods introduced to Newtonian optics, especially after the mid-1990s, can be adopted and implemented to the cosmological light propagation problems. 

In order to achieve this, we considered the geodesic action for two neighboring null curves simultaneously. The equivalent, geodesic deviation action is then obtained via the method introduced in  \citep{Vines:2014} by Synge's world function. Under the thin ray bundle approximation, we considered the terms up to quadratic order only and obtained the action with respect to the tetrad components of the geodesic deviation variables. This allowed us to define a 4-dimensional reduced phase space and a corresponding quadratic Hamiltonian function. Note that in the conventional approach of cosmology, the optical phase space is 6-dimensional. It is composed of three spatial components of spacetime \textit{coordinates} and the 3-momentum of the photon. In our reduced phase space, however, it is the Sachs basis components of the \textit{deviation vector} and its total derivative with respect to the affine parameter are the ones that compose the phase space vector. This makes our approach Machian in nature. In addition, having an optical phase space composed of the tetrad components of the variables directly links the ray bundle evolution to the observables in question. 

Quadratic Hamiltonians are encountered in many areas of physics, whose associated flows are given by linear Hamilton-Jacobi equations. The advantage of quadratic polynomials is that the Lie operators constructed through them have matrix representations. Moreover, as the exponential maps of Hamiltonian matrices are symplectic matrices, the corresponding phase space transformations are then represented by symplectic transfer matrices. In our case, the symplectic ray bundle transfer matrix is written as a $4\times 4$ block matrix composed of $2\times 2$ submatrices $\mathbf{A}, \mathbf{B}, \mathbf{C}$ and $\mathbf{D}$ as it is usually referred to as ABCD matrices in the Newtonian optics community. 

Etherington's distance reciprocity, which follows from the first order geodesic deviation equation, is then shown to hold for light propagation from any initial and final points, not just when the initial point is a vertex. This follows from the symplectic conditions imposed on the ray bundle transfer matrix and holds within any spacetime.

Symplectomorphisms on our reduced phase space are linear canonical transformations. Generating function of such ABCD canonical transformations were discussed in the literature previously. We showed that it corresponds to our quadratic geodesic deviation action and we wrote it in the form of matrix inner products of initial and transformed phase space coordinates. We also introduced a phase space distribution function and the corresponding Liouville equation.

In the end, we proposed some potential applications of our formalism to show its full power. In particular, we suggested that: 
\begin{itemize}
\item[(i)] The reduced phase space averaging, that leads us to a null bundle average, is a relevant tool to estimate the averaged observables in our past null cone. It can be used to average the full set of scalar spin field equations in the Newman-Penrose formalism to study the average effect of the full set of the Einstein equations.
\item[(ii)] Iwasawa factorization, or other types of decomposition techniques of symplectic matrices, can be used to identify the unique elements of an optical system which, in our case, is the spacetime. Then light  {ray} propagation in any spacetime between any initial and final points can be factored into its thin lens, pure magnifier and fractional Fourier transformer components  {for the case of thin bundles}.
\item[(iii)] Wavization of an observed thin null bundle is possible due to the relation between the symplectic and metaplectic groups. Following the method outlined in  \citep{Moshinsky:1971} and used in Newtonian optics, we showed that the kernel of the diffraction integral is given by the generating function of the underlying linear canonical transformation. In our case it is exactly equal to our quadratic deviation action. 
\item[(iv)] Evolution of the polarization states of the CMB can be investigated via the recent techniques developed in Newtonian optics. Those include the evolution of the generalized Stokes parameters as the beam propagates through an ABCD system and constructed by an optical Wigner distribution function. Those methods can then be adopted to investigate the polarization within a generic, nonperturbative spacetime geometry.
\item[(v)] The technical difficulty of estimating the ray bundle transfer matrix via  ordered exponentials can be overcome by applying an additional canonical transformation on an extended phase space to autonomize the system. If such a transformation exists for a given phase space then the corresponding evolution can be obtained by Lie transformations. Moreover, the stability analysis and phase portraits for observed null bundles can then be determined for a given spacetime.
\end{itemize}

 {Note that even though current paper focuses on the propagation of light rays, the methods introduced here are, in principle, applicable for the propagation of gravitational waves as well. In the case of gravitational radiation, the corresponding wave vector follows the null geodesics of a background spacetime \citep{Sachs:1961, Ohanian:1974, Dai:2016}. This is especially relevant for the studies of lensing and focusing of gravitational waves. In the event that the wavelength is longer than the lensing object's radius, diffraction effects become important which provides information about both the source and the lensing object \citep{Nakamura:1997, Lai:2018}. Those effects can in fact be investigated within the wavization procedure we introduced in Section (\ref{Wavization of a ray bundle}).}

Here we acknowledge another work that was introduced into the literature during the preparation of this manuscript. In \citep{Grasso:2019}, the authors obtain a ray bundle transfer (or Wronskian) matrix via bilocal operators, which is similar by construction to our formulation. Their transfer matrix, including higher degrees of freedom, might indeed lead one to the generalization of the method and potential applications presented here.  

As a final remark we note that when discussions of possible deviations from cosmological distance reciprocity is put forward \citep{Bassett:2003, Bassett:2003zw, Lazkoz:2007, Rasanen:2015}, it is usually stated that there are three possible explanations for such a deviation (if it exists) \citep{Yang:2013, Ellis:2013}: (i) light does not propagate on a Riemannian geometry; (ii) the geometric optics approximation is broken, i.e.,  the light does not follow null geodesics; (iii) the number of photons is not conserved throughout the propagation due to the coupling of axions, gravitons, etc. Here we state that  {a potential breakdown of distance reciprocity necessarily follows from a deviation of symplecticity of the system. Note that this complies with all of the possible explanations of reciprocity breakdown scenarios given above.}

 {In conclusion, we stress that in general relativity; (i) physically meaningful quantities are obtained through geodesic deviation, (ii) the observables are measured on our local frames. Additionally, in the classical theory, symplectic symmetries are closely related to the observables of a given system. In classic statistical mechanics and in quantum
mechanics, for example, symplectic symmetries play a major role and the observables are those that are measured by the ones that are already in their local frames. Therefore, it is not surprising that our symplectic symmetries emerge once we introduce local projections of the null geodesic deviation vector on our observational screen.  Note that all of those different constructions in physics can be linked via phase space methods for linear systems.}
\ack
The author is grateful to David L. Wiltshire, Miko\l{}aj Korzy\'nski, Volker Perlick and Thomas Buchert for valuable comments and discussions.  {I also thank to the referee for pointing out the potential application of the methods introduced here to lensing of gravitational waves.} This work was supported by the grant GACR 14-37086G of Albert Einstein Center for Gravitation and Astrophysics of the Czech Science Foundation. This work is also a part of a project that has received funding from the European Research Council (ERC) under the European Union’s Horizon 2020 research and innovation programme (grant agreement ERC advanced grant 740021–ARTHUS, PI: {Thomas Buchert}).

\section*{Appendix: Derivation of Equation \texorpdfstring{set~(\ref{eq:trans_evol})}{}}\label{myAppendix}
\appendix
\setcounter{section}{1}
\subsection{Useful expressions}
Here we present certain expressions that will be relevant for our derivation. Note that we are using the numerator layout notation in our derivations. 
\subsubsection{\texorpdfstring{$\boldsymbol{\xi'}$ and $\boldsymbol{\dot{\xi'}}$}{}}
Let us consider the following ray bundle transfer
\begin{eqnarray}\label{eq:tr_xiprimetoxi}
\left[\begin{array}{c}
 \boldsymbol{\xi} \\
 \boldsymbol{\dot{\xi}}
\end{array}\right]= 
\left[
\begin{array}{c|c}
\mathbf{A} & \, \, \mathbf{B} \\
\hline
\mathbf{C} & \, \, \mathbf{D}
\end{array}
\right]_{(v,v')}
\left[\begin{array}{c}
 \boldsymbol{\xi'} \\
 \boldsymbol{\dot{\xi'}}
\end{array}\right].
\end{eqnarray}
Then, through (\ref{eq:Inv_symplectic}) we have
\begin{eqnarray}
\left[\begin{array}{c}
 \boldsymbol{\xi'} \\
 \boldsymbol{\dot{\xi'}}
\end{array}\right]= 
\left[
\begin{array}{c|c}
\mathbf{D^{\intercal}} & \, \, -\mathbf{B^{\intercal}} \\
\hline
-\mathbf{C^{\intercal}} & \, \, \mathbf{A^{\intercal}}
\end{array}
\right]_{(v,v')}
\left[\begin{array}{c}
 \boldsymbol{\xi} \\
 \boldsymbol{\dot{\xi}}
\end{array}\right],
\end{eqnarray}
and together with the evolution equations (\ref{eq:four_sets})
\begin{eqnarray}
\boldsymbol{\xi'}&=\mathbf{\dot{B}}^{\intercal}\boldsymbol{\xi}-\mathbf{B}^{\intercal}\boldsymbol{\dot{\xi}},\label{eq:xiprime}\qquad \qquad
\boldsymbol{\dot{\xi'}}&=-\mathbf{\dot{A}}^{\intercal}\boldsymbol{\xi}+\mathbf{A^{\intercal}}\boldsymbol{\dot{\xi}}\label{eq:dotxiprime}.
\end{eqnarray}
\subsubsection{Symmetry of \texorpdfstring{$\,\mathbf{\dot{B}}\mathbf{B}^{\rm{-1}}$}{} and \texorpdfstring{$\mathbf{B}^{\rm{-1}}\mathbf{A}$}{}}
When we substitute the evolution equations (\ref{eq:four_sets}) into equation set (\ref{eq:Block_inv}), first two lines follow as
\begin{eqnarray}
\mathbf{\dot{B}}^{\intercal}&=\left(\mathbf{A}-\mathbf{B}\mathbf{\dot{B}}^{-1}\mathbf{\dot{A}}\right)^{-1}\label{eq:dotBTrans}\\
-\mathbf{B}^{\intercal}&=-\left(\mathbf{A}-\mathbf{B}\mathbf{\dot{B}}^{-1}\mathbf{\dot{A}}\right)^{-1}\mathbf{B}\mathbf{\dot{B}}^{-1}\label{eq:minusBTrans}.
\end{eqnarray}
Substitution of (\ref{eq:dotBTrans}) into (\ref{eq:minusBTrans}) gives
$
(\mathbf{B}\mathbf{\dot{B}}^{\rm{-1}})^{\intercal}=(\mathbf{B}\mathbf{\dot{B}}^{\rm{-1}}),
$
and taking the inverse gives
\begin{eqnarray}\label{eq:symm_BdotBinv}
\mathbf{\dot{B}}\mathbf{B}^{\rm{-1}}=\left(\mathbf{\dot{B}}\mathbf{B}^{\rm{-1}}\right)^{\intercal}.
\end{eqnarray}
Now, consider the first two lines of the equation set (\ref{eq:recip_matrices}) and the evolution equations (\ref{eq:four_sets}). Then we have
\begin{eqnarray}
\mathbf{A}\left(v,v'\right)&=\mathbf{\dot{B}}^{\intercal}\left(v',v\right),\qquad \qquad
\mathbf{B}\left(v,v'\right)&=-\mathbf{B}^{\intercal}\left(v',v\right).\nonumber
\end{eqnarray}
With these and (\ref{eq:symm_BdotBinv}) we have the symmetry of the product $\mathbf{B}^{\rm{-1}}\mathbf{A}$, i.e.,
\begin{eqnarray}\label{eq:symm_BinvA}
\mathbf{B}^{\rm{-1}}\mathbf{A}=\left(\mathbf{B}^{\rm{-1}}\mathbf{A}\right)^{\intercal}.
\end{eqnarray}
\subsection{Derivation of \texorpdfstring{$\boldsymbol{\dot{\xi}}={\partial S}/{\partial \boldsymbol{\xi}}$}{}}
Let us substitute the evolution equations (\ref{eq:four_sets}) into the generating function given in (\ref{eq:gen_fun}) in order to get
\begin{eqnarray}\label{eq:gen_fun_evol}
S=\frac{1}{2}\left[\boldsymbol{\xi}^{\intercal}\left(\mathbf{\dot{B}}\mathbf{B}^{\rm{-1}}\right)^{\intercal}\boldsymbol{\xi}\right]&-&\boldsymbol{\xi}^{\intercal}\left(\mathbf{B}^{\rm{-1}}\right)^{\intercal}\boldsymbol{\xi'}
+\frac{1}{2}\left[\boldsymbol{\xi'}^{\intercal}\left(\mathbf{B}^{\rm{-1}}\mathbf{A}\right)^{\intercal}\boldsymbol{\xi'}\right].
\end{eqnarray}
Then, due to (\ref{eq:symm_BdotBinv}), we have the following 
\begin{eqnarray}
\frac{\partial S}{\partial \boldsymbol{\xi}}=\frac{1}{2}\left(2\boldsymbol{\xi}^{\intercal}\mathbf{\dot{B}}\mathbf{B}^{\rm{-1}}\right)-\boldsymbol{\xi'}^{\intercal}\mathbf{B}^{\rm{-1}}.
\end{eqnarray}
Now, let us substitute (\ref{eq:xiprime}) into above so that we get
\begin{eqnarray}
\frac{\partial S}{\partial \boldsymbol{\xi}}=\boldsymbol{\xi}^{\intercal}\mathbf{\dot{B}}\mathbf{B}^{\rm{-1}}-\boldsymbol{\xi}^{\intercal}\mathbf{\dot{B}}\mathbf{B}^{\rm{-1}}+\boldsymbol{\dot{\xi}}^{\intercal}\mathbf{B}\mathbf{B}^{\rm{-1}}=\boldsymbol{\dot{\xi}}^{\intercal}.
\end{eqnarray}
\subsection{Derivation of \texorpdfstring{$\boldsymbol{\dot{\xi'}}=-{\partial S}/{\partial \boldsymbol{\xi'}}$}{}}
Taking the derivative of $S$ in  the form of (\ref{eq:gen_fun_evol}) with respect to $\boldsymbol{\xi'}$ this time gives
\begin{eqnarray}
\frac{\partial S}{\partial \boldsymbol{\xi'}}=-\boldsymbol{\xi}^{\intercal}\left(\mathbf{B}^{\rm{-1}}\right)^{\intercal}+\frac{1}{2}\left[2\boldsymbol{\xi'}\left(\mathbf{B}^{\rm{-1}}\mathbf{A}\right)^{\intercal}\right],
\end{eqnarray}
due to (\ref{eq:symm_BinvA}). Now let us substitute $\boldsymbol{\xi}=\mathbf{A}\boldsymbol{\xi'}+\mathbf{B}\boldsymbol{\dot{\xi'}}$ into above following the transfer (\ref{eq:tr_xiprimetoxi}) to get
\begin{eqnarray}
\frac{\partial S}{\partial \boldsymbol{\xi'}}&=&-\boldsymbol{\xi'}^{\intercal}\mathbf{A}^{\intercal}\left(\mathbf{B}^{\rm{-1}}\right)^{\intercal}-\boldsymbol{\dot{\xi'}}^{\intercal}\mathbf{B}^{\intercal}\left(\mathbf{B}^{\rm{-1}}\right)^{\intercal}+\boldsymbol{\xi'}\left(\mathbf{B}^{\rm{-1}}\mathbf{A}\right)^{\intercal}\nonumber
\end{eqnarray}
Then,
\begin{eqnarray}
\frac{\partial S}{\partial \boldsymbol{\xi'}}=-\boldsymbol{\dot{\xi'}}^{\intercal}.
\end{eqnarray}
\subsection{Derivation of \texorpdfstring{${\partial S}/{\partial v}+H=0$}{}}
Once we substitute (\ref{eq:xiprime}) into (\ref{eq:gen_fun_evol}), the partial derivative of $S$ with respect to the affine parameter, $v$, follows as
\begin{eqnarray}\label{eq:partialSpartialv}
\frac{\partial S}{\partial v}=\frac{1}{2}\boldsymbol{\xi}^{\intercal}\One\,\boldsymbol{\xi}+\frac{1}{2}\boldsymbol{\xi}^{\intercal}\Two\,\boldsymbol{\dot{\xi}}+\frac{1}{2}\boldsymbol{\dot{\xi}}^{\intercal}\Three\,\boldsymbol{\xi}+\frac{1}{2}\boldsymbol{\dot{\xi}}^{\intercal}\Four\,\boldsymbol{\dot{\xi}},
\end{eqnarray}
in which
\begin{eqnarray}
\One =-\left[\left(\mathbf{B}^{\rm{-1}}\right)^{\intercal}\right]^{\boldsymbol{\cdot}}\mathbf{\dot{B}}^{\intercal}+\left(\mathbf{B}^{\rm{-1}}\right)^{\intercal}(\mathbf{\ddot{B}})^{\intercal}
+\mathbf{\dot{B}}\left[\left(\mathbf{B}^{\rm{-1}}\mathbf{A}\right)^{\intercal}\right]^{\boldsymbol{\cdot}}\mathbf{\dot{B}}^{\intercal}.
\end{eqnarray}
Once we substitute the evolution equation $\mathbf{\ddot{B}}=\mathbf{\dot{D}}=\boldsymbol{\mathcal{R}}\mathbf{B}$ given in equation set (\ref{eq:four_sets}) into the second term on the right hand side of the above, we obtain
\begin{eqnarray}
\One &=&\boldsymbol{\mathcal{R}}^{\intercal} {+}\Big\{\mathbf{\dot{B}}\left[-(\mathbf{B}^{\rm{-1}})^{\boldsymbol{\cdot}}+\left(\mathbf{B}^{\rm{-1}}\mathbf{A}\right)^{\boldsymbol{\cdot}}\mathbf{\dot{B}}^{\intercal}\right]\Big\}^{\intercal}\nonumber\\
&=&\boldsymbol{\mathcal{R}}^{\intercal} {+}\Big\{\mathbf{\dot{B}}\left[(\mathbf{B}^{\rm{-1}})^{\boldsymbol{\cdot}}\left(-\mathbf{I_2}+\mathbf{A}\mathbf{\dot{B}}^{\intercal}\right)+\mathbf{B}^{\rm{-1}}\mathbf{\dot{A}}\mathbf{\dot{B}}^{\intercal}\right]\Big\}^{\intercal}.\nonumber
\end{eqnarray}
Due to the second line of the symplecticity conditions (\ref{eq:symm_matrices}), we have
\begin{eqnarray}
\One &=& \boldsymbol{\mathcal{R}}^{\intercal} {+}\Big\{\mathbf{\dot{B}}\left[(\mathbf{B}^{\rm{-1}})^{\boldsymbol{\cdot}}\mathbf{B}\mathbf{\dot{A}}^{\intercal}+\mathbf{B}^{\rm{-1}}\mathbf{\dot{A}}\mathbf{\dot{B}}^{\intercal}\right]\Big\}^{\intercal}\nonumber \\
&=& \boldsymbol{\mathcal{R}}^{\intercal} {+}\Big\{\mathbf{\dot{B}}\left[-\mathbf{B}^{\rm{-1}}\mathbf{\dot{B}}\mathbf{\dot{A}}^{\intercal}+\mathbf{B}^{\rm{-1}}\mathbf{\dot{A}}\mathbf{\dot{B}}^{\intercal}\right]\Big\}^{\intercal},
\end{eqnarray}
where we use the generic derivation
\begin{eqnarray}\label{eq:Binvdot}
(\mathbf{B}^{\rm{-1}})^{\boldsymbol{\cdot}}=-\mathbf{B}^{\rm{-1}}\mathbf{\dot{B}}\mathbf{B}^{\rm{-1}}
\end{eqnarray}
to obtain the second line. Then we have
\begin{eqnarray}\label{eq:One_Result}
\One = \boldsymbol{\mathcal{R}}^{\intercal},
\end{eqnarray}
due to the symmetry of $\mathbf{\dot{A}}\mathbf{\dot{B}}^{\intercal}$ that is given as a symplectic condition in the first line of equation set (\ref{eq:symm_matrices}).

The second term in equation (\ref{eq:partialSpartialv}) follows as
\begin{eqnarray}
\Two&=&2\left[(\mathbf{B}^{\rm{-1}})^{\intercal}\right]^{\boldsymbol{\cdot}}\mathbf{B}^{\intercal}-\mathbf{\dot{B}}\left[\mathbf{A}^{\intercal}(\mathbf{B}^{\rm{-1}})^{\intercal}\right]^{\boldsymbol{\cdot}}\mathbf{B}^{\intercal}\nonumber \\
&=&2\left[(\mathbf{B}^{\rm{-1}})^{\boldsymbol{\cdot}}\right]^{\intercal}\mathbf{B}^{\intercal}-\mathbf{\dot{B}}\left[\mathbf{B}^{\rm{-1}}\mathbf{A}\right]^{\boldsymbol{\cdot}}\mathbf{B}^{\intercal}\nonumber \\
&=&2\left[(\mathbf{B}^{\rm{-1}})^{\boldsymbol{\cdot}}\right]^{\intercal}\mathbf{B}^{\intercal}+\mathbf{\dot{B}}\mathbf{B}^{\rm{-1}}\left(\mathbf{\dot{B}}\mathbf{A}^{\intercal}-\mathbf{\dot{A}}\mathbf{B}^{\intercal}\right)\nonumber \\
&=&2\left(-\mathbf{B}^{\rm{-1}}\mathbf{\dot{B}}\mathbf{B}^{\rm{-1}}\right)^{\intercal}\mathbf{B}^{\intercal}+\mathbf{\dot{B}}\mathbf{B}^{\rm{-1}}\nonumber \\
&=&-2(\mathbf{B}^{\rm{-1}})^{\intercal}\mathbf{\dot{B}}^{\intercal}+\mathbf{\dot{B}}\mathbf{B}^{\rm{-1}},
\end{eqnarray}
in which we make use of (\ref{eq:symm_BinvA}) and (\ref{eq:Binvdot}) to obtain the third line. The fourth line follows from the symplectic condition given in the second line of (\ref{eq:symm_matrices}) and (\ref{eq:Binvdot}). Hence we have
\begin{eqnarray}\label{eq:Two_Result}
\Two=-\mathbf{\dot{B}}\mathbf{B}^{\rm{-1}},
\end{eqnarray}
due to (\ref{eq:symm_BdotBinv}).

The third term in (\ref{eq:partialSpartialv}) is as follows.
\begin{eqnarray}
\Three&=&-\mathbf{B}\left[\mathbf{A}^{\intercal}(\mathbf{B}^{\rm{-1}})^{\intercal}\right]^{\boldsymbol{\cdot}}\mathbf{\dot{B}}^{\intercal}\nonumber\\
&=&\mathbf{B}\left(\mathbf{B}^{\rm{-1}}\mathbf{\dot{B}}\mathbf{B}^{\rm{-1}}\right)\mathbf{A}\mathbf{\dot{B}}^{\intercal}-\mathbf{\dot{A}}\mathbf{\dot{B}}^{\intercal}\nonumber \\
&=&\mathbf{\dot{B}}\mathbf{B}^{\rm{-1}}\left(\mathbf{I_2}+\mathbf{B}\mathbf{\dot{A}}^{\intercal}\right)-\mathbf{\dot{A}}\mathbf{\dot{B}}^{\intercal},
\end{eqnarray}
in which we use (\ref{eq:symm_BinvA}) to obtain the second line and use the second line of the symplectic conditions (\ref{eq:symm_matrices}) to obtain the third.
Due to the symmetry of $\mathbf{\dot{A}}\mathbf{\dot{B}}^{\intercal}$ given in (\ref{eq:symm_matrices}), we have
\begin{eqnarray}\label{eq:Three_Result}
\Three=\mathbf{\dot{B}}\mathbf{B}^{\rm{-1}}.
\end{eqnarray}

The fourth term of (\ref{eq:partialSpartialv}) is
\begin{eqnarray}
\Four&=&\mathbf{B}\left[\mathbf{A}^{\intercal}(\mathbf{B}^{\rm{-1}})^{\intercal}\right]^{\boldsymbol{\cdot}} {\mathbf{B}^{\intercal}}\nonumber\\
&=&\mathbf{B}\mathbf{\dot{A}}^{\intercal}-\mathbf{B}\mathbf{A}^{\intercal}\mathbf{\dot{B}}\mathbf{B}^{\rm{-1}}\nonumber\\
&=&\left(\mathbf{A}\mathbf{\dot{B}}^{\intercal}-\mathbf{I_2}\right)-\mathbf{B}\mathbf{A}^{\intercal}\mathbf{\dot{B}}\mathbf{B}^{\rm{-1}}\nonumber \\
&=&\mathbf{A}\mathbf{\dot{B}}^{\intercal}-\mathbf{I_2}-\mathbf{A}\mathbf{B}^{\intercal}(\mathbf{B}^{\rm{-1}})^{\intercal}\mathbf{\dot{B}}^{\intercal},
\end{eqnarray}
where we use (\ref{eq:Binvdot}) and (\ref{eq:symm_BdotBinv}) to obtain the second line. The third line follows from the second line of (\ref{eq:symm_matrices}). The fourth line is obtained by making use of the first line of the symplectic conditions (\ref{eq:symm_matrices}) and (\ref{eq:symm_BdotBinv}). 
Thus,
\begin{eqnarray}\label{eq:Four_Result}
\Four=-\mathbf{I_2}.
\end{eqnarray}

Now, let us substitute the results given in (\ref{eq:One_Result}), (\ref{eq:Two_Result}), (\ref{eq:Three_Result}) and (\ref{eq:Four_Result}) into (\ref{eq:partialSpartialv}). Then we have
\begin{eqnarray}
\frac{\partial S}{\partial v}=-\left[\frac{1}{2}\left(\boldsymbol{\dot{\xi}},\boldsymbol{\dot{\xi}}\right)-\frac{1}{2}\left(\boldsymbol{\mathcal{R}}\boldsymbol{\xi},\boldsymbol{\xi}\right)\right]=-H,
\end{eqnarray}
in which $H$ is our reduced quadratic Hamiltonian given in (\ref{eq:Red_Hamiltonian}). 
\bibliographystyle{iopart-num}
                                                                                                                                                                                                                                                                                                                                                                                                                                                                                                                                                                                                                                                                                                                                                                                                                                                                                                                                                                                                                                                                                                                                                                                                                                                                                                                                                                                                                                                                                                                                                                                                                                                                                                                                                                                                                                                                                                                                                                                                                                                                                                                                                                                                                                                                                                                                                                                                                                                                                                                                                                                                                                                                                                                                                                                                                                                                                                                                                                                                                                                                                                                                                                                                                                                                                                                                                                                                                                                                                                                                                                                                                                                                                                                                                                                                                                         \bibliography{references}
                                                                                                                                                                                                                                                                                                                                                                                                                                                                                                                                                                                                                                                                                                                                                                                                                                                                                                                                                                                                                                                                                                                                                                                                                                                                                                                                                                                                                                                                                                                                                                                                                                                                                                                                                                                                                                                                                                                                                                                                                                                                                                                                                                                                                                                                                                                                                                                                                                                                                                                                                                                                                                                                                                                                                                                                                                                                                                                                                                                                                                                                                                                                                                                                                                                                                                                                                                                                                                                                                                                                                                                                                                                                                                                                                                                                                                         
\end{document}